\documentclass[a4paper,11pt]{article}
\pdfoutput=1
\usepackage{jheppub}
\usepackage[mathscr]{eucal}
\usepackage{hhline}
\usepackage{psfrag}
\usepackage{mathrsfs} 
\usepackage{hyperref}
\usepackage{bm}
\usepackage{array}
\usepackage{tikz}
\usepackage{subcaption}
\usepackage{textcomp}
\usepackage{hyperref}
\usepackage{mathrsfs} 
\usepackage{enumitem}
\usepackage{caption}
\usepackage{epic}
\usepackage{eepic}
\usepackage{gensymb}
\DeclareMathAlphabet{\mathpzc}{OT1}{pzc}{m}{it}

\makeatletter
\@addtoreset{equation}{section}
\makeatother

\def\bea{\begin{eqnarray}}
\def\eea{\end{eqnarray}}
\def\be{\begin{equation}}
\def\ee{\end{equation}}

\def\be{\begin{equation}}
\def\ee{\end{equation}}
\def\bdm{\begin{displaymath}}
\def\edm{\end{displaymath}}
\def\bea{\begin{eqnarray}}
\def\eea{\end{eqnarray}}

\def\ri{{\rm i}}
\def\half{\textstyle\frac{1}{2}}

\def\XXint#1#2#3{{\setbox0=\hbox{$#1{#2#3}{\int}$}
    \vcenter{\hbox{$#2#3$}}\kern-.5\wd0}}

\newcommand{\rd}{\mbox{d}}
\newcommand{\re}{\mbox{e}}

\DeclareMathAlphabet{\mathpzc}{OT1}{pzc}{m}{it}

\usetikzlibrary{shapes.misc,decorations.pathmorphing,calc}
\tikzset{
  branch cut/.style={
    decorate,decoration=snake,
    to path={
      (\tikztostart) -- (\tikztotarget) \tikztonodes
    }
  }
}

\notoc
\title{On the scaling behaviour of the alternating spin chain}
\author[a]{Vladimir V. Bazhanov,}
\author[a,b]{Gleb A.  Kotousov,}
\author[a]{Sergii M. Koval,}
\author[b,c]{Sergei  L. Lukyanov}

\affiliation[a]{Department of Theoretical Physics,
         Research School of Physics and Engineering,\\
    Australian National University, Canberra, ACT 2601, Australia}
\affiliation[b]{NHETC, Department of Physics and Astronomy,
     Rutgers University,\\
     Piscataway, NJ 08855-0849, USA}
\affiliation[c]{Kharkevich Institute for Information Transmission Problems,\\
Moscow, 127994, Russia}

\emailAdd{vladimir.bazhanov@anu.edu.au}
\emailAdd{kotoousov@physics.rutgers.edu}
\emailAdd{sergii.koval@anu.edu.au}
\emailAdd{sergei@physics.rutgers.edu}

\abstract{In this note we report  the results of  our study of a 1D integrable spin chain whose critical behaviour
is governed by a CFT possessing a continuous spectrum
of scaling dimensions. It is argued that the computation of the density of Bethe states 
of the continuous theory can be reduced to the calculation of the connection coefficients for
a certain class of differential equations whose monodromy properties are similar to those of
the conventional confluent hypergeometric  equation. The finite size corrections to the scaling are also discussed.
}

\arxivnumber{1903.05033 }

\begin{document}
\captionsetup[figure]{labelfont={small},labelformat={default},labelsep=period,name={Fig.\!}}
\captionsetup[table]{labelfont={small},labelformat={default},labelsep=period,name={Tab.\!}}

\maketitle
\vskip 4cm 
\pagebreak

Going back to  Kadanoff's  block spin transformation, the concept of the Renormalization Group
(RG) is usually illustrated  by   means of  a finite statistical  lattice system that provides a regularization of 
the Euclidean path integral.   
Within the Hamiltonian picture, attempts to introduce  the scale transformation
 for a finite lattice system meet immediate difficulties. Of course, 
 since the Hilbert space is not isomorphic for different lattice sizes,
  the scale transformation only makes sense for the low energy 
part of the spectrum. 
It is
clear how to assign the size dependence for the ground state  or, for that matter, the lowest energy
states in the disjoint sectors of the Hilbert space. 
However 
 forming  individual  RG flows trajectories 
 for low energy stationary states that are 
densely distributed does not seem to be a trivial task.
One dimensional quantum spin chains provide an
ideal laboratory for studying this problem.
\bigskip

In the case of a critical  spin chain  subject to (quasi) periodic  boundary conditions,
conformal invariance predicts a so-called tower structure \cite{Cardy:1986ie}  for the excitation energy $\Delta E$ over the
ground state for the  low-energy part of the spectrum:
\bea\label{eq1}
\Delta E=\frac{2\pi v_{\rm F}}{L}\ \big(d_s+{N}+\bar{N}\big)+o(L^{-1})\ .
\eea
Here $d_s$ are the scaling dimensions of the conformal primary; $N$, $\bar{N}$ are non-negative 
integers describing the excitations;  and $v_{\rm F}$ is the Fermi velocity.
In many cases the finite size corrections, denoted by $o(L^{-1})$, turn 
out to be small even for a lattice size $L$ that is not too large.
Thus, despite the large degeneracies, the tower structure is often useful for
identifying the scaling behaviour of the stationary states on the finite lattice
provided that the spectrum of the scaling dimensions $\{d_s\}$ is discreet.
In the presence of a continuous component in the set $\{d_s\}$,
the practical use of eq.\,\eqref{eq1} becomes problematic.
\bigskip

Critical spin chain systems exhibiting a continuous spectrum of scaling dimensions are of
considerable interest in many aspects, including the quantization of 2D non-linear sigma models
on non-compact spaces and their applications to the description of condensed matter systems with
disorder. In the work  \cite{Jacobsen:2005xz} the remarkable observation was made that 
the critical behaviour of the alternating spin chain
originally introduced in \cite{Baxter:1971,Baxter} is described 
by a CFT possessing a continuous spectrum
of scaling dimensions. The model turns out to be an integrable system,
which makes a detailed study of its RG flow possible. This was the subject of
the papers \cite{Ikhlef:2008zz,Ikhlef:2011ay,Frahm:2012eb,Candu:2013fva,Frahm:2013cma}, 
where some important results were obtained.
In this note we present a summary of our study of the RG flow in the
alternating spin chain. A detailed analysis and derivations will be given elsewhere.  
\bigskip

\begin{figure}[t]
\centering
\scalebox{1}{
\includegraphics[height=2.4cm]{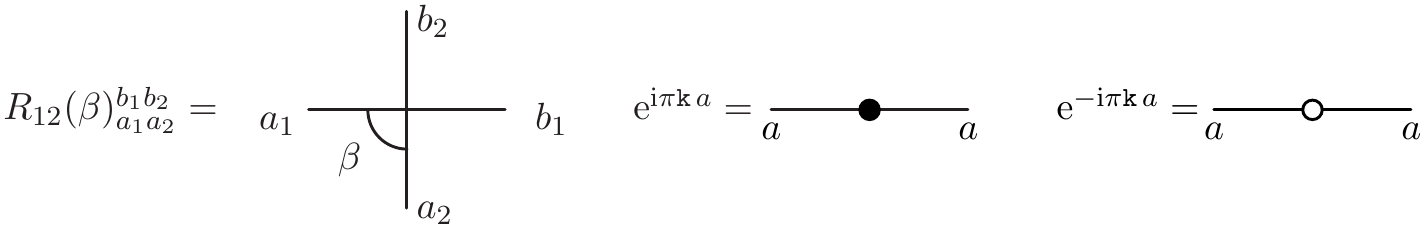}}
\caption{\small Graphical representation of the $R$-matrix \eqref{rmat} and the 
boundary twists $\re^{\pm \ri \pi {\mathtt k}\sigma_z}$.
The edge indices $a,a_1,a_2,b_1,b_2$ take two values $\pm1$.} 
\label{pic1}
\end{figure}

\begin{figure}[ht!]
\centering
\scalebox{0.9}{
\includegraphics[height=3.9cm]{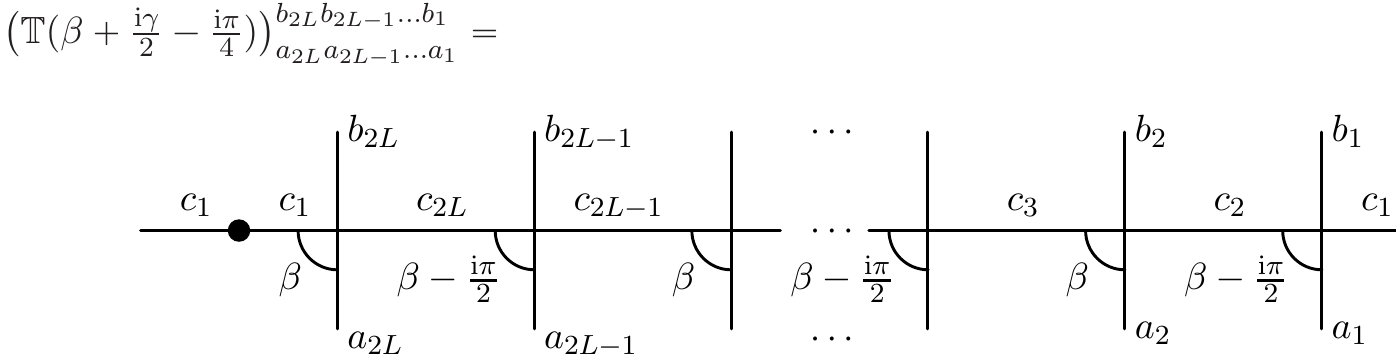}}
\caption{\small Graphical representation of the transfer matrix \eqref{tmat}. The summation over the spin indices assigned to internal edges is assumed.} 
\label{pic2}
\end{figure}

The subject of our interest is a spin-$\frac{1}{2}$ chain of length $2L$ governed by the Hamiltonian
\bea\label{iasusay}
{\mathbb H}&=&\frac{1}{\sin(2\gamma)}\
\sum_{m=1}^{2L}\,\Big(\,2\sin^2(\gamma)\, \sigma^z_m\,\sigma^z_{m+1}- (\sigma^x_m\,\sigma^x_{m+2}+\sigma^y_m\,\sigma^y_{m+2}+
\sigma^z_m\,\sigma^z_{m+2})\\[0.2cm]
&+& 
\ri\, (-1)^m\sin(\gamma)
(\sigma^x_m\,\sigma^y_{m+1}
-\sigma^y_m\,\sigma^x_{m+1})(\sigma^z_{m-1}-\sigma^z_{m+2})
\,\Big)+2L\,\cot(2\gamma)\ .
\nonumber
\eea
In order to lift degeneracies in the energy spectrum as much as possible,
instead of the periodic spin chain
we will consider  quasi periodic boundary conditions
\bea\label{sisisaisu}
\sigma^{\pm}_{2L+m}=\re^{\pm 2\ri\pi{\tt k}}\ \sigma^{\pm}_{m}\ ,\ \ \ \ \ \ \ \sigma^{z}_{2L+m}=\sigma^{z}_{m}\ \ \ \ \ \ \ \ \ 
\big(\sigma^\pm\equiv\half\, (\sigma^x\pm\ri\,\sigma^y)\,\big)\ ,
\eea
involving the parameter ${\tt k}$ lying within the ``first Brillouin zone''
$$-\half < {\tt k}\leq\half \ .$$
The Hamiltonian \eqref{iasusay} is connected to an alternating 6-vertex model. Its $R$-matrix, $R_{12}(\beta)$, is an operator depending on the spectral parameter $\beta$ and acting in the product of two vector spaces
${\mathbb C}^2\otimes{\mathbb C}^2$. We denote its matrix elements as 
$R_{12}(\beta)_{a_1a_2}^{b_1b_2}$, where the indices take two values $\pm1$.
The indices $a_1,b_1$ and $a_2,b_2$ refer to the first and second spaces, respectively. 
There are only six non-zero matrix elements of  $R_{12}(\beta)$, which are given by
\begin{eqnarray}
R_{12}(\beta)^{++}_{++}=&R_{12}(\beta)^{--}_{--}=1\,,\qquad 
R_{12}(\beta)^{+-}_{+-}=R_{12}(\beta)^{-+}_{-+}=\displaystyle
\frac{\sinh(\beta)}{\sinh(\beta+\ri\gamma)}\,\nonumber\\[.3cm]
&R_{12}(\beta)^{+-}_{-+}=R_{12}(\beta)^{-+}_{+-}
=\displaystyle\frac{\sinh(\ri\gamma)}{\sinh(\beta+\ri\gamma)}\,.\label{rmat}
\end{eqnarray}
Using the transfer matrix 
\begin{eqnarray}
\big({\mathbb T}(\beta+{\textstyle\frac{\ri\gamma}{2}
  -\frac{\ri\pi}{4}})\big)_{a_{2L}a_{2L-1}\ldots a_1}^{b_{2L}b_{2L-1}\ldots b_1}&=&
\sum_{c_1,c_2,\ldots,c_{2L}}\,\re^{\ri \pi {\mathtt k} c_1}\ 
R(\beta)_{c_1\,\,\, \,a_{2L}}^{c_{2L}b_{2L}}\,
R(\beta-{\textstyle\frac{\ri\pi}{2}})_{c_{2L}\,\,\,\,\,\,\, a_{2L-1}}^{c_{2L-1} b_{2L-1}}\cdots\nonumber\\[.3cm]
&&\qquad \qquad \qquad \qquad \cdots\ \,R(\beta)_{c_3a_{2}}^{c_{2} b_{2}}\,
R(\beta-{\textstyle\frac{\ri\pi}{2}})_{c_{2}a_{1}}^{c_{1} b_{1}}\ ,
\label{tmat}\end{eqnarray}
the Hamiltonian  \eqref{iasusay} can be expressed as
\begin{equation}\label{Heq1}
{\mathbb H}=-\ri\partial_\beta\,\log\big({\mathbb T}(\beta)\,{\mathbb T}(\beta+\tfrac{\ri\pi}{2})\big)\big|_{\beta=\frac{\ri\gamma}{2}-\frac{\ri\pi}{4}}\,.
\end{equation}
It convenient to represent the $R$-matrix graphically as in fig.\,\ref{pic1}.
Then, with these conventions the transfer matrix is represented as in fig.\,\ref{pic2}.
\bigskip

The system, thus defined, can be studied using the Bethe Ansatz (BA) approach and the corresponding 
equations read
explicitly as \cite{Lieb:1967,Baxter:1971}
\bea\label{aisasausa}
\bigg(\frac{\cosh(2\beta_j+\ri\gamma)}{\cosh(2\beta_j-\ri\gamma)}
\bigg)^L=-\re^{-2\ri\pi{\tt k}}\ \prod_{m=1}^M\frac{\sinh(\beta_j-\beta_m+\ri\gamma)}{\sinh(\beta_j-\beta_m-\ri\gamma)}\ .
\eea
For a chain of given length $2L$ every solution of the BA equations 
corresponds to an  eigenstate of the Hamiltonian \eqref{iasusay} with energy
\be\label{Eeq1}
E=-\sum_{j=1}^M\ \frac{4\sin(2\gamma)}{\cosh(4\beta_j)+\cos(2\gamma)}\ .
\ee
The number of Bethe roots, $M$,  is related to the total spin,
$\frac{1}{2}\,\sum_j\,\sigma^z_j$, which turns out to be a conserved quantity 
for the chain
\be\label{Meq1}
M=L-S^z\ .
\ee
\smallskip

Along with the $U(1)$-symmetry generated by the total spin operator, the system
admits the global ${\cal C}{\cal P}$-invariance that acts on the spins as
\bea
{\cal C}{\cal P}\,\sigma^{{{\pm}}}_m\,
{\cal C}{\cal P}=\sigma^{{{\mp}}}_{2L+1-m}\, , \ \ \ \ \ \ \ 
{\cal C}{\cal P}\,\sigma^{{{z}}}_m\,
{\cal C}{\cal P}=-\sigma^{{{z}}}_{2L+1-m}\ \ \ \ \ \ (m=1,\ldots, 2L)\ .
\eea
The latter intertwines the sectors with $+S^z$ and $-S^z$ so that,
without loss of generality, we will focus our attention on the case with
\be
M\le L\ .\nonumber
\ee
Another global symmetry is ${\cal C}{\cal P}{\cal T}$, which acts inside each sector
with given spin $S^z$. It manifests itself in the BA equations as the invariance
of the system \eqref{aisasausa} w.r.t. complex conjugation:
\be\label{CPTeq1}
{\cal C}{\cal P}{\cal T}\  : \ \ \ \ \ \ \ \ \ \beta_j \mapsto \beta_j^*\ \ \ \ \ \ \ \ \ ({\rm mod}\ \ri\pi)\, .
\ee
The spin chain possesses yet another $\mathbb{Z}_2$-symmetry. 
The explicit formula for the operator, ${\cal D} \, :  \ {\cal D}^2=1$,  that generates it
 is not important for our purposes and can be found in the papers \cite{Saleur:1990vd,Jacobsen:2005xz,Ikhlef:2011ay} (it is denoted
 by $C$ therein).
This $\mathbb{Z}_2$-symmetry corresponds to the invariance of eq.\,\eqref{aisasausa}
w.r.t. the transformation
\be\label{Deq2}
\ \ \ \ {\cal D}\ : \ \ \ \ \ \ \ \ \  \beta_j \mapsto \beta_j+\tfrac{\ri\pi}{2}\ \ \ \ \ \ \ ({\rm mod}\ \ri\pi)\, .
\ee

\begin{figure}
\centering
\begin{subfigure}[b]{0.49\textwidth}
\scalebox{0.9}{
\begin{tikzpicture}
\draw [->,thick] (-3.8,0) -- (3.8,0);
\draw [->,thick] (0,-3) -- (0,3);
\node at (3.5,2.6) {$\beta$};
\draw  (3.51,2.63) circle [radius=0.3];
\draw[thick,branch cut] (0,2.5) to (0,1);
\draw[thick,branch cut] (0,-2.5) to (0,-1);
\draw[black,fill=black] (0,1) circle (.5ex);
\draw[black,fill=black] (0,-1) circle (.5ex);
\draw[black,fill=black] (0,2.5) circle (.5ex);
\draw[black,fill=black] (0,-2.5) circle (.5ex);
\node at (1.2,1) {$+\tfrac{1}{2}\big(\frac{\pi}{2}-\gamma\big)$};
\node at (1.2,2.5) {$+\tfrac{1}{2}\big(\frac{\pi}{2}+\gamma\big)$};
\node at (-1.2,-1) {$-\tfrac{1}{2}\big(\frac{\pi}{2}-\gamma\big)$};
\node at (-1.2,-2.5) {$-\tfrac{1}{2}\big(\frac{\pi}{2}+\gamma\big)$};
\end{tikzpicture}
}
\end{subfigure}
\begin{subfigure}[b]{0.49\textwidth}
\scalebox{0.9}{
\begin{tikzpicture}
\draw [->,thick] (-3.8,0) -- (3.8,0);
\draw [->,thick] (0,-3) -- (0,3);
\node at (2.7,2.7) {$\beta$};
\draw  (2.71,2.73) circle [radius=0.3];
\draw[thick,branch cut] (+3.8,1.5) to (0,1.5);
\draw[thick,branch cut] (-3.8,-1.5) to (0,-1.5);
\draw[black,fill=black] (0,1.5) circle (.5ex);
\draw[black,fill=black] (0,-1.5) circle (.5ex);
\node at (-0.6,1.5) {$+\gamma$};
\node at (0.8,-1.6) {$-\gamma$};
\end{tikzpicture}
}
\end{subfigure}
\caption{\small
The complex $\beta$-plane displaying the branch cuts for the functions
$P$  (left panel) and $\Theta$ (right panel) that are closest to the origin.
Subsequent cuts are obtained by shifting this picture by $\ri\pi N$ with integer $N$.
\label{branch1}}
\end{figure}
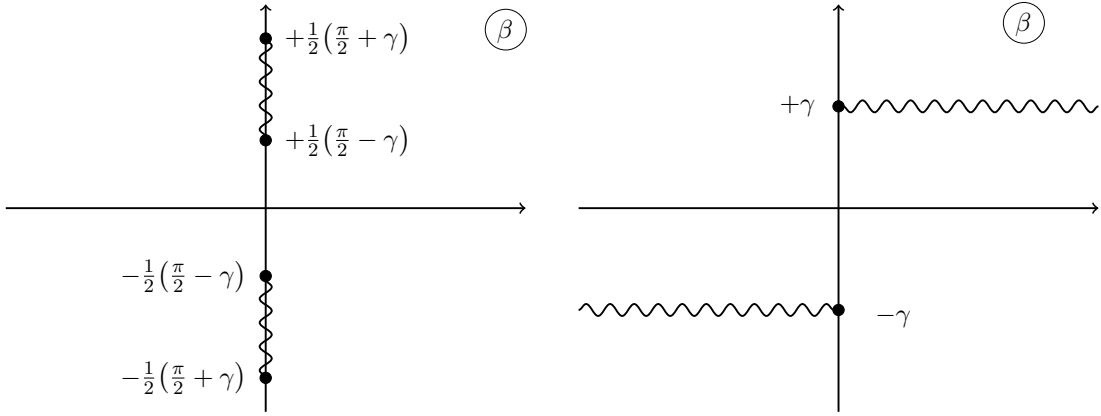

\begin{figure}
\centering
\scalebox{0.82}{
\begin{tikzpicture}
\node at (0,0) {\includegraphics[width=0.52\textwidth]{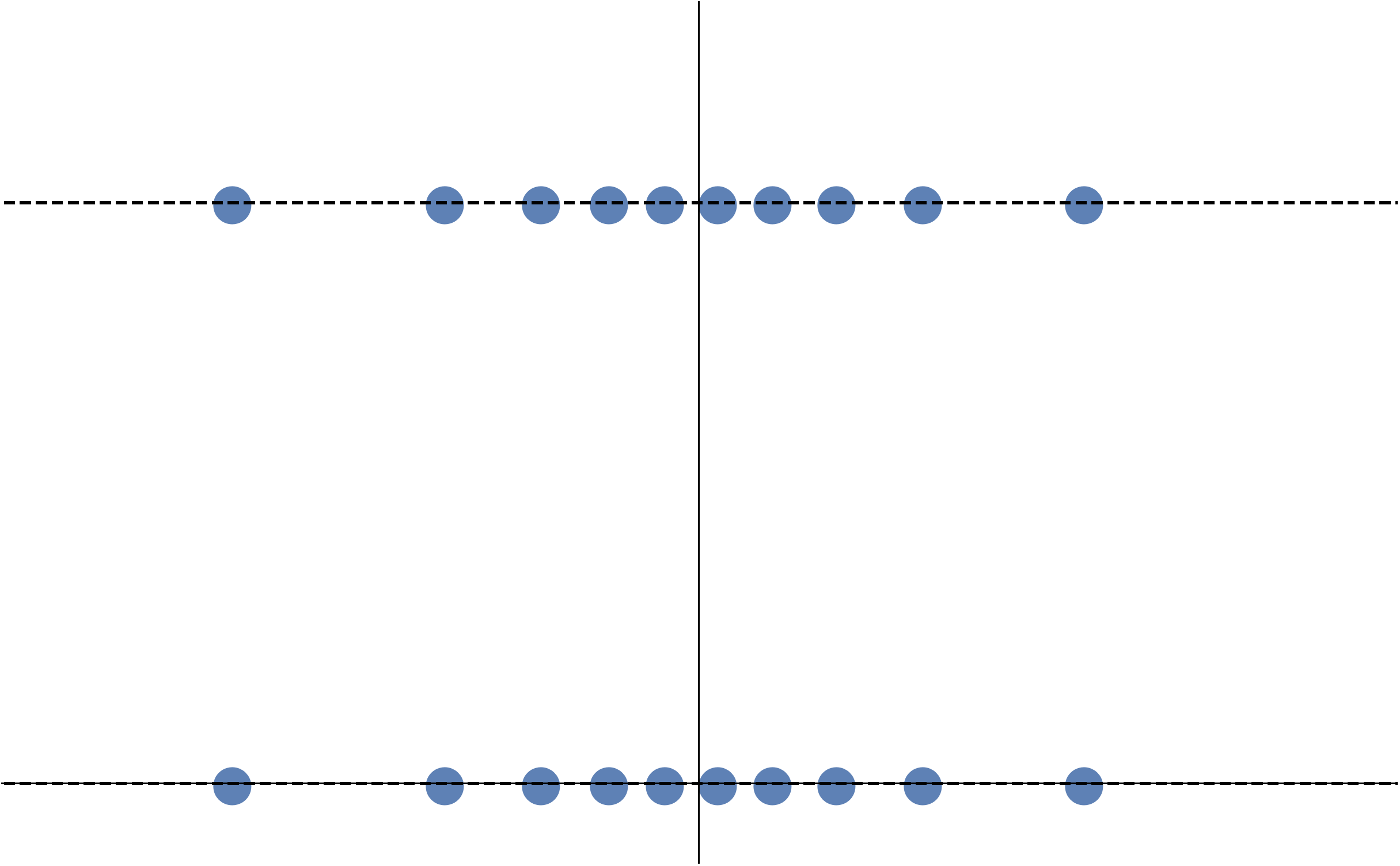}};
\node at (9.3,0.1) {\includegraphics[width=0.52\textwidth]{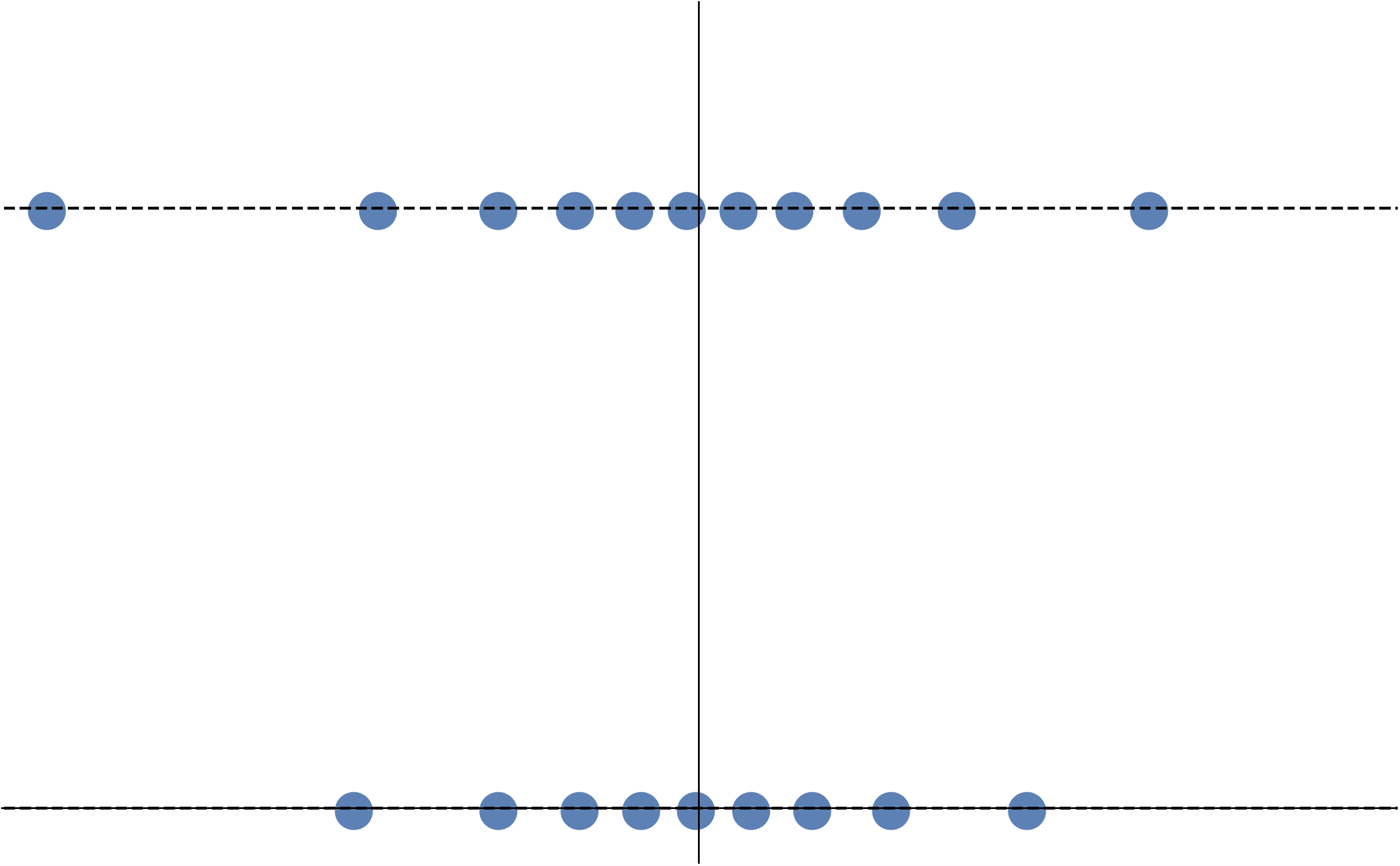}};
\node at (3,0.5) {$\scriptstyle\beta$};
\draw  (3.02,0.54) circle [radius=0.2];
\node at (12.3,0.5) {$\scriptstyle\beta$};
\draw  (12.32,0.54) circle [radius=0.2];
\node at (3.3,1.65) {$\scriptstyle\Im m(\beta)=\frac{\pi}{2}$};
\node at (3.3,-1.7) {$\scriptstyle\Im m(\beta)=0$};
\node at (13.2,1.65) {$\scriptstyle\Im m(\beta)=\frac{\pi}{2}$};
\node at (13.2,-1.7) {$\scriptstyle\Im m(\beta)=0$};
\node at (-2.8,1.65) {\small $ 1$};
\node at (-1.6,1.65) {\small $ 2$};
\node at (-1,1.65) {\small $ 3$};
\node at (-0.1,1.6) { \large $ \ldots$};
\node at (1.1,1.65) {\small $ 9$};
\node at (2,1.65) {\small $ 10$};
\node at (-2.8,-1.65) {\small $ 11$};
\node at (-1.6,-1.65) {\small $ 12$};
\node at (-0.95,-1.65) {\small $ 13$};
\node at (1.1,-1.65) {\small $ 19$};
\node at (2,-1.65) {\small $ 20$};
\node at (-0.1,-1.7) {\large  $ \ldots$};
\node at (5.6,1.65) {\small $1$};
\node at (7.35,1.65) {\small $2$};
\node at (8,1.65) {\small $3$};
\node at (8.8,1.6) {\large $\ldots$};
\node at (10.6,1.65) {\small $10$};
\node at (11.6,1.65) {\small $11$};
\node at (7.1,-1.65) {\small $12$};
\node at (8,-1.65) {\small $13$};
\node at (10.2,-1.65) {\small $19$};
\node at (11,-1.65) {\small $20$};
\node at (8.8,-1.7) {\large $\ldots$};
\end{tikzpicture}
}
\caption{\small The pattern of Bethe roots in the complex plane for $L=20$,
$\gamma=\frac{\pi}{5}$, ${\tt k} = \frac{1}{10}$ and with $S^z=0$.
The integers beside each point show the labeling of the
roots $\beta_j$. The corresponding Bethe numbers 
are given by eq.\,\eqref{BN1}.
The left panel corresponds to the ground state for which the number of roots on the
lines $\Im m(\beta)=0$ and $\frac{\pi}{2}$ are equal ($m=0$); whereas on the
right panel there are two more roots on the upper line than the lower one, i.e.,
$m=2$.
\label{BAplot1}
}
\end{figure}
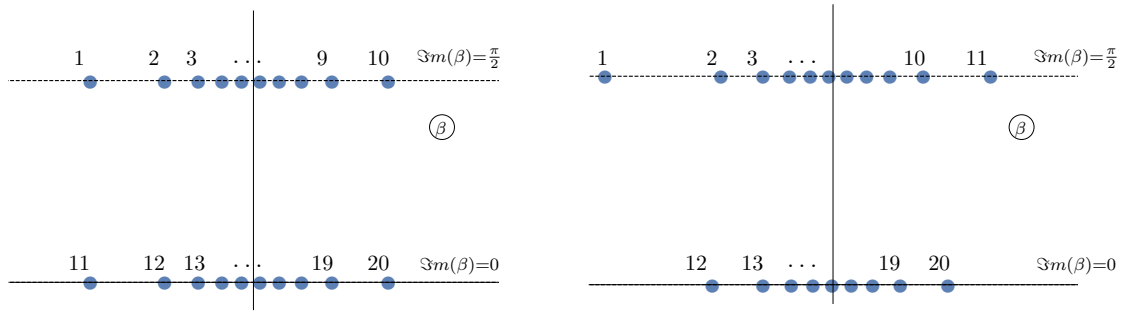

The assigning of a scale dependence to the low energy stationary states is greatly facilitated by
the existence of the BA equations and
can be done along the following line. First of all, eq.\,\eqref{aisasausa} should be re-written 
in logarithmic form:
\bea\label{klkwqnmsd}
L P(\beta_j)=2\pi\, I_j-2\pi{\tt k}-\sum_{m=1}^M \Theta(\beta_j-\beta_m)\ ,
\eea
where
\bea\label{func1}
P(\beta)=\frac{1}{\ri}\ \log\bigg[\frac{\cosh(\ri\gamma+2\beta)}{\cosh(\ri\gamma-2\beta)}\bigg] \ , \qquad 
\Theta(\beta)=\frac{1}{\ri}\ \log\bigg[\frac{\sinh(\ri\gamma-\beta)}{\sinh(\ri\gamma+\beta)}\bigg]\ ,
\eea
while $I_j$ are the so-called Bethe numbers which are integers or half-integers for $M$ odd or even respectively. 
In order to define the Bethe numbers unambiguously one should specify the branches of the multivalued
functions \eqref{func1}. We do this by imposing the conditions 
$$
P(0)=\Theta(0)=0
$$
and choosing the system of branch cuts as shown in fig.\,\ref{branch1}.
It is important to mention that our analysis is restricted to the spin chain 
with the parameter 
\be
0<\gamma<\frac{\pi}{2}\ .
\ee
In this case the set of Bethe numbers corresponding to the lowest energy state
 in the sector with spin $S^z$ is given by
\be\label{BN1}
I_j=-\tfrac{1}{2}\,(M+1)+j\ \ \ \ \ \ \ \ (j=1,2,\ldots, M)\,,
\ee
valid for any $L$ (the labeling of the Bethe roots is explained in the caption in fig.\,\ref{BAplot1}).
It should be emphasized that the $\{I_j\}$ do not uniquely specify the solution
of the BA equations. For example, in the case of the vacuum in the sector with given $S^z$ 
the Bethe roots are distributed along the lines 
$\Im m(\beta)=0,\frac{\pi}{2} \ ({\rm mod} \, \pi)$. For even $M$ the vacuum is non-degenerate
and the Bethe roots are equally distributed along these lines while for $M$ odd, the vacuum is two
fold degenerate corresponding to an excess of one of the roots with $\Im m(\beta_j)=0$ or
$\frac{\pi}{2}$. Notice that in the latter case, the vacua are related by the 
$\mathbb{Z}_2$-transformation \eqref{Deq2}. 
In fact, the BA equations \eqref{klkwqnmsd} with Bethe numbers as in \eqref{BN1}
admit solutions such that the difference between the number of roots with
$\Im m(\beta_j)=\frac{\pi}{2}$ and 
$\Im m(\beta_j)=0$ is equal to $m$, where (for an illustration see fig.\,\ref{BAplot1})
\be\label{mformula1}
m=0,\,\pm2,\,\pm4,\ldots \ \ \ {\rm for} \ \ M{\rm \ even} \ , \ \ \ \ \ \ \ \ \ \ 
m=\pm1,\,\pm3,\ldots \ \ \ {\rm for} \ \ M{\rm\  odd} \ .
\ee
In the rest of this note, to simplify the discussion, we make the technical assumption that $M$ is even.

\bigskip

The numerical solution of  the BA equations \eqref{klkwqnmsd} not only requires the
specification of the Bethe numbers, but also a proper initial approximation
for the positions of the Bethe roots. The latter turns out to be the most difficult part
of the numerical procedure. In our studies we approached the problem in the following way.
Starting with a spin chain for relatively small $L$ ($L\le 12$ in our analysis), we performed the
numerical diagonalization of the Hamiltonian. The latter is part of a family of commuting operators,
of which a prominent r$\hat{{\rm o}}$le is played by the so-called $Q$-operator.
Together with the eigenvalues of the Hamiltonian, we computed the corresponding 
eigenvalues of $Q$, which turn out to be polynomials w.r.t. the variable
$\re^{-2\beta}$. The zeroes of this polynomial coincide with $\re^{-2\beta_j}$, where
the $\beta_j$ solve the BA equations \eqref{aisasausa}.\footnote{%
In fact there are two commuting $Q$-operators $Q_\pm$ and eq.\,\eqref{aisasausa}
with $M\le L$ is satisfied by the zeroes of the operator $Q_+$. For the case $S^z<0$ one
should consider the zeroes of $Q_-$, which obey the equations similar to \eqref{aisasausa}
with $M=L+S^z$ and the twist parameter ${\tt k}\mapsto -{\tt k}$.
}
Thus for each 
Bethe state, i.e., the stationary state which is simultaneously an eigenvector of 
the $Q$-operator, we were able to find the set $\{\beta_j\}$ and,
using eq.\,\eqref{klkwqnmsd} as a definition of the Bethe numbers,
 the corresponding
$\{I_j\}$. 
The numerical values of the Bethe roots for some $L$ can be used to construct
an initial approximation for the solution of the BA equations \eqref{klkwqnmsd} 
with $L$ increased to $L+2$ for which the qualitative pattern of the roots 
remains the same. We found this to be a highly effective procedure for 
determining the RG flow of an individual Bethe state.

\bigskip

\begin{figure}[t]
\centering
\scalebox{0.9}{
\includegraphics[height=4.9cm]{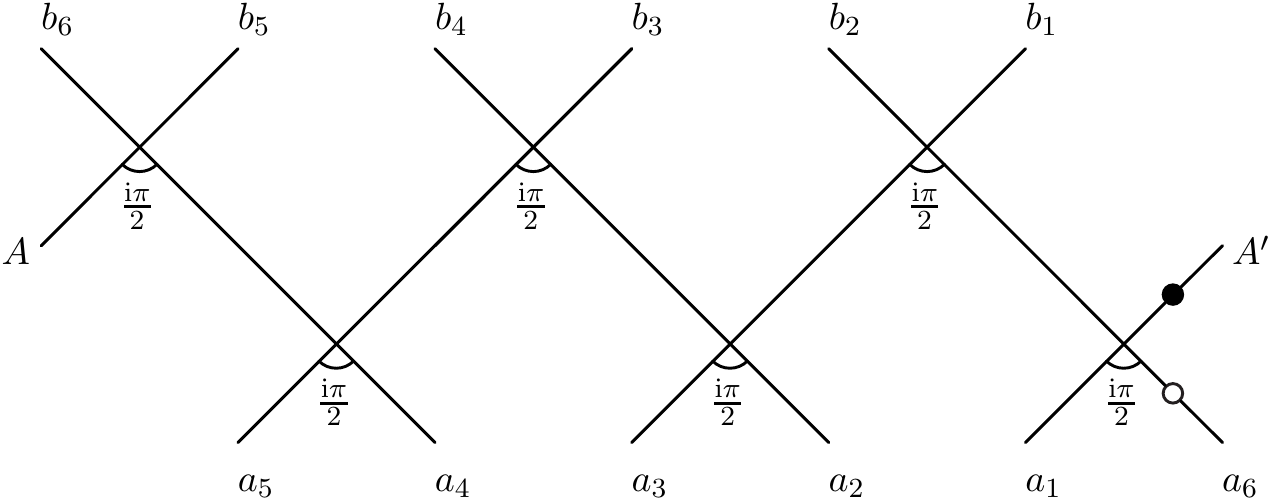} }
\vspace{0.2cm}
\caption{\small A graphical representation of the 
matrix elements
$({\mathbb B})_{a_{2L}a_{2L-1}\ldots a_1}^{b_{2L}\,b_{2L-1}\ldots b_1}$
of the quasi-shift operator \eqref{qshift} for the chain of length $2L=6$.
Summation over the spin indices assigned to internal edges is assumed. Note that for (quasi) periodic 
boundary conditions, $A$ and $A'$ should be identified.} 
\label{pic3}
\end{figure}

With the above method it is possible to study the scale dependence of 
various observables. Along with the energy $E(L)$ computed by means of eq.\,\eqref{Eeq1}
 we also focused on
the eigenvalue of the so-called quasi-shift operator. This is an important observable that
commutes with the Hamiltonian and was introduced in \cite{Ikhlef:2011ay}. 
It is defined using the transfer matrix \eqref{tmat} as 
\begin{equation}
{\mathbb B}={\mathbb T}\big(\textstyle{\frac{\ri}{2}(\gamma+\frac{\pi}{2})}
\big)
\big[{\mathbb T}\big(\textstyle{\frac{\ri}{2}(\gamma-\frac{\pi}{2})}\big)\big]^{-1}\,.\label{qshift}
\end{equation}
Interestingly, this operator can be viewed as a transfer matrix of a homogeneous 
(not alternating) six vertex model on the $45\degree$-rotated regular square lattice
 with quasi-periodic boundary conditions. Indeed, with the conventions of fig.\,\ref{pic1}, 
 the operator $\mathbb{B}$ can be represented as in fig.\,\ref{pic3}.
In terms of the Bethe roots, the eigenvalues of $\mathbb{B}$ are given by
\bea\label{Beq1}
B(L)=\prod_{j=1}^M\frac{\cosh(2\beta_j)-\sin(\gamma)}{\cosh(2\beta_j)+\sin(\gamma)}\  .
\eea
Finally another useful characteristic of the flow is the product 
\be\label{Pieq1}
\Pi(L)=\prod_{j=1}^M\,\re^{4\beta_j}\ ,
\ee
which can be considered as the eigenvalue of an operator that appears
naturally in the large-$\beta$ expansion of the $Q$-operator.
\bigskip

Before discussing the general features of $E(L)$, $B(L)$ and $\Pi(L)$,
it is  useful to get a feel of them for some particular classes of Bethe states.
To this end, we restrict ourselves for now to the states with the simplest pattern of
Bethe roots, illustrated in fig.\,\ref{BAplot1}, that were previously mentioned.
The eigenvalue $B(L)$ for these states was already discussed in the literature  \cite{Ikhlef:2011ay,Candu:2013fva,Frahm:2013cma}.
It was pointed out that for large $L$ the quantity
\be\label{seq1}
s(L)= \frac{n}{4\pi}\,\log (B)
\ee
with
\be
n=\frac{\pi}{\gamma}-2>0 \nonumber
\ee
behaves as
\be\label{seq2}
s(L)\asymp\frac{\pi m}{4\log(L)}\,.
\ee
Here $m$ is the difference between the number of roots with
$\Im m (\beta_j)=\frac{\pi}{2}$ and $\Im m (\beta_j)=0$ \eqref{mformula1}.
The formula \eqref{seq2} resembles the quantization condition of a quantum mechanical particle in a potential well of length
$\propto \log(L)$. 
It turns out, that as in usual quantum mechanics, a more accurate quantization condition is achieved by taking into account
the phase shift that the particle picks up in the vicinity of the turning points.
The results of our analysis yields the following quantization condition
\bea\label{quantC1}
8s\  \log\bigg(\frac{2L\Gamma(\frac{3}{2}+\frac{1}{n})}{\sqrt{\pi}\Gamma(1+\frac{1}{n})}\bigg)
+\delta(s)-2\pi m=O\big((\log L)^{-\infty}\big)\ .
\eea
The phase shift entering the above equation is explicitly given by the formula
\bea\label{phaseeq1}
\delta(s)=\frac{16s}{n}\ \log(2)-2\ri\ \log\bigg[2^{4\ri s}
\ \frac{\Gamma(\frac{1}{2}+p-{\ri s})\Gamma(\frac{1}{2}+{\bar p}-{\ri s})}{\Gamma(\frac{1}{2}+p+{\ri s})\Gamma(\frac{1}{2}+{\bar p}+{\ri s})}\, \bigg]
\eea
with
\be\label{peq1}
p=\tfrac{1}{2}\,\big(S^z+{\tt k}\,(n+2)\big)\, , \qquad \bar{p}=\tfrac{1}{2}\,\big(S^z-{\tt k}\,(n+2)\big)\, .
\ee
The quantization condition holds true up to power law corrections in $L$, 
as indicated by the r.h.s. of eq.\,\eqref{quantC1}. The quality of the approximation
is illustrated in tab.\,\ref{tab1}.
Also,
it should be noted that the phase shift \eqref{phaseeq1} was essentially guessed in ref.\cite{Ikhlef:2011ay}.

\begin{table}
\centering
\begin{tabular}{|c|c|c|c|}
\hline
& & & \\[-0.4cm]
$L$ & $s$ from eqs.\,\eqref{Beq1}\,\eqref{seq1} & $s$ from eq.\,\eqref{quantC1} & $\big({\rm r.h.s.\ of}\ \eqref{quantC1}\big)\times L^2$ \\[0.1cm]
\hline
& & & \\[-0.45cm]
10 & 1.102443315 & 1.138091292 & -75.8926  \\[0.05cm]
\hline
& & & \\[-0.45cm]
20 & 0.897548147 & 0.904491191 & -75.6208  \\[0.05cm]
\hline
& & & \\[-0.45cm]
50 & 0.712588922 & 0.713466560 & -76.4795 \\[0.05cm]
\hline
& & & \\[-0.45cm]
100 & 0.615519194 & 0.615709067 & -76.9693 \\[0.05cm]
\hline
& & & \\[-0.45cm]
200 & 0.541653004 & 0.541694888 & -77.3589 \\[0.05cm]
\hline
& & & \\[-0.45cm]
400 & 0.483644642 & 0.483654019 & -77.6930 \\[0.05cm]
\hline
& & & \\[-0.45cm]
800 & 0.436891428& 0.436893552 & -77.9932 \\[0.05cm]
\hline
\end{tabular}
\caption{\small In the second column the values of $s$ were obtained by solving the 
BA equations for increasing $L$ with the parameters
 $\gamma=\frac{\pi}{6}$ ($n=4$), ${\tt k}=\frac{1}{5}$, $S^z=4$, $m=4$ and then using
eqs.\,\eqref{Beq1}\,\eqref{seq1}. These are compared with predictions coming from
the quantization condition \eqref{quantC1}. In the last column
$s$ obtained from the BA equations is substituted into the l.h.s. of 
eq.\,\eqref{quantC1} to measure the corrections 
$O\big((\log L)^{-\infty}\big)$. It indicates that for $n=4$
these corrections
are of the order $L^{-2+\epsilon}$ with $\epsilon\to +0$.
\label{tab1}
}
\end{table}

\bigskip

The large-$L$ asymptotic of the eigenvalue $\Pi(L)$ is expressed in terms of $s=s(L)$ \eqref{seq1}.
The relation, again valid up to powers of $L$, explicitly 
reads as
\be\label{omegaeq1}
\Pi(L)=\Omega\ \bigg[\frac{2 L\,\Gamma\big(\frac{3}{2}+\frac{1}{n}\big)}{\sqrt{\pi}\,\Gamma\big(1+\frac{1}{n}\big)}\bigg]^{\frac{2n(\bar{p}-p)}{n+2}}\,
\big(1+O\big((\log L)^{-\infty}\big)\big)
\ee
with
\be\label{omegaeq2}
\Omega= 2^{2({\bar p}-p)}\ \ 
(n+2)^{\frac{4(\bar{p}-p)}{n+2}} \ \Bigg[\frac{\Gamma\big(1+\frac{2\bar{p}}{n+2}\big)\,\Gamma(1+2p)}
{\Gamma\big(1+\frac{2p}{n+2}\big)\,\Gamma(1+2\bar{p})}\Bigg]^2\ 
\ \frac{\Gamma\big(\frac{1}{2}+\bar{p}+\ri s\big)\,\Gamma\big(\frac{1}{2}+\bar{p}-\ri s\big)}
{\Gamma\big(\frac{1}{2}+p+\ri s\big)\,\Gamma\big(\frac{1}{2}+p-\ri s\big)}\,\ .
\ee
Of course, the above formula can be applied only for the specific class of Bethe states we have been currently focused on.
Finally the excitation energy of these states above the ground state turns out to be
\bea\label{asdasd}
\Delta E(L) = \frac{2\pi v_F}{L}\bigg(\frac{p^2+{\bar p}^2}{n+2}+\frac{2s^2}{n}\bigg) +o\big(L^{-1}\big)\,,
\eea
where $s=s(L)$ and the Fermi velocity reads as  
\be\label{Vfermi}
v_{\rm F}=\frac{2(n+2)}{n}\ .
\ee
\smallskip

The formula \eqref{asdasd} suggests that the Bethe states under consideration 
flow to conformal primaries as $L\to \infty$,
with scaling dimensions given by the term in the parenthesis in the r.h.s.
However it follows from eq.\,\eqref{seq2} that the value of
 $s(L)$ tends to zero for large $L$ if $m$ is left unchanged. A non-trivial scaling limit
 can be achieved by increasing $|m|$ simultaneously with $L$ in such a way that 
 $s$ is kept fixed.  Then the scaling dimensions pick up a component 
 labeled by this continuous parameter. Thus the CFT underlying
 the critical behaviour of the alternating spin chain possesses a continuous spectrum
 of scaling dimensions.
\bigskip

Spectroscopy of the low energy excitations of the alternating spin chain reveals 
another class of states which, as $L\to \infty$, flows to  conformal primaries characterized 
by the pair of
conformal dimensions $(\bar{\Delta},{\Delta})$ with
\be\label{cdeq1}
\Delta=\frac{p^2}{n+2}+\frac{s^2}{n}\, , \ \ \ \ \ {\bar \Delta}=\frac{{\bar p}^2}{n+2}+\frac{s^2}{n}
\ee
and
\be\label{cdeq2}
p=\tfrac{1}{2}\,\big(S^z+({\tt k}+{\tt w})\,(n+2)\big)\, , \qquad \bar{p}=\tfrac{1}{2}\,\big(S^z-({\tt k}+{\tt w})\,(n+2)\big)\, .
\ee
The last formula is analogous to eq.\,\eqref{peq1} but with ${\tt k}$ shifted by the integer
 ${\tt w}=\pm 1,\pm 2\ldots\ $. Since the Hamiltonian is a periodic function of ${\tt k}$ 
 (see eqs.\,\eqref{iasusay},\,\eqref{sisisaisu}) the integer ${\tt w}$ enumerates
 the different bands of the spectrum. 
 We will refer to the corresponding states as winding states. 
\bigskip

\begin{figure}
\centering
\scalebox{0.8}{
\begin{tikzpicture}
\node at (0,0) {\includegraphics[width = 0.55\textwidth]{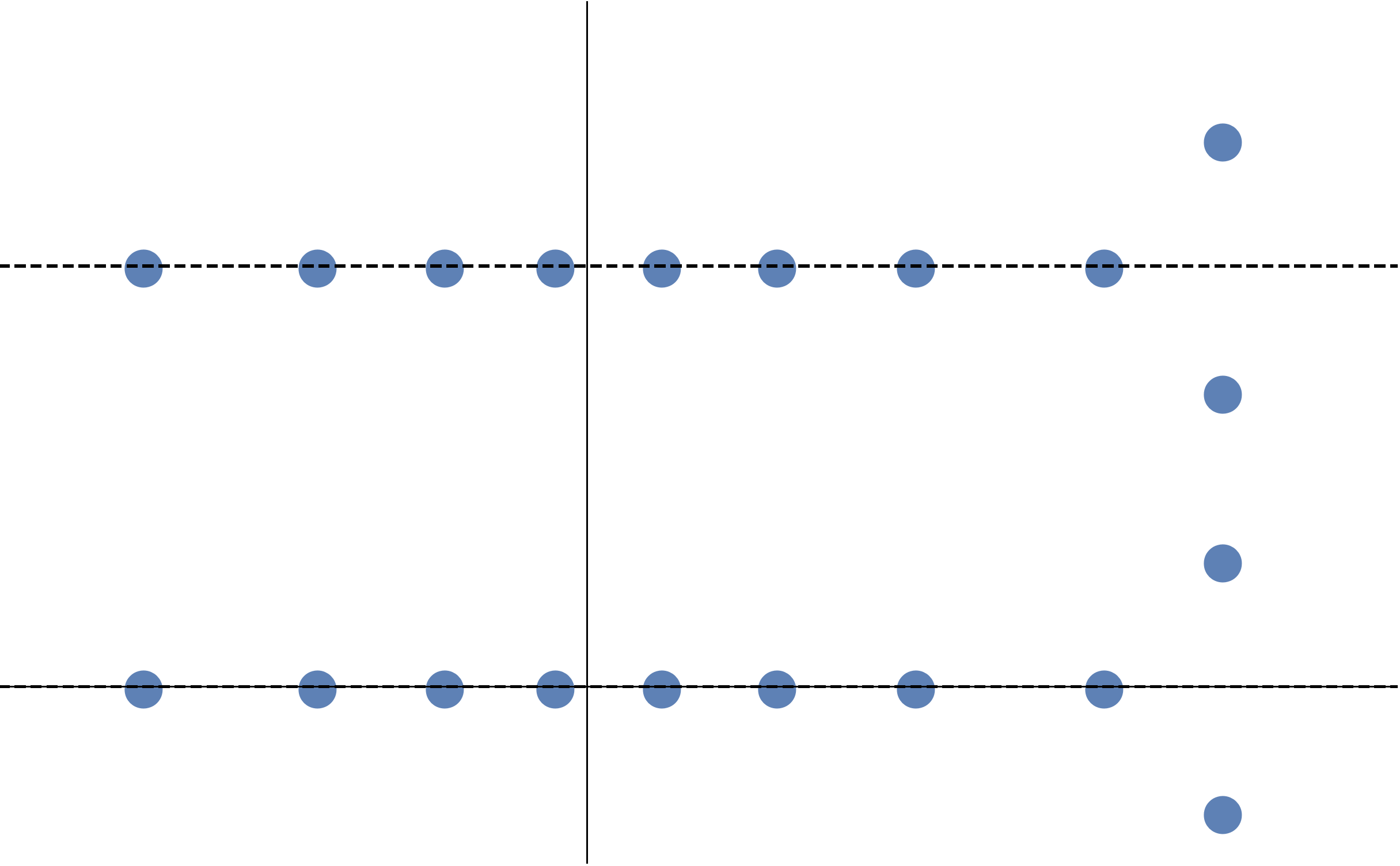}};
\node at (-3.5,1.35) {\small $1$};
\node at (-2.5,1.35) {\small $2$};
\node at (-1.5,1.3) {\large $\ldots$};
\node at (2.3,1.35) {\small $8$};
\node at (3.5,1.8) {\small $9$};
\node at (3.6,0.25) {\small $10$};
\node at (-3.5,-1.15) {\small $11$};
\node at (-2.5,-1.15) {\small $12$};
\node at (-1.5,-1.2) {\large $\ldots$};
\node at (2.3,-1.15) {\small $18$};
\node at (3.6,-0.75) {\small $19$};
\node at (3.6,-2.3) {\small $20$};
\end{tikzpicture}
}
\caption{\small The pattern of Bethe roots for the winding state with ${\tt w}=-1$, $S^z=0$
and $L=20$.
With the labeling as shown, the corresponding Bethe numbers are given by 
the reference distribution \eqref{BN2}. 
The value of the parameters for the plot were taken to be
$\gamma=\frac{\pi}{5}$ and ${\tt k}=\frac{1}{30}$. 
\label{fig30}
}
\end{figure}
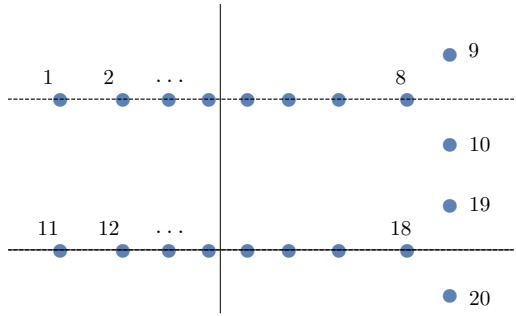
It is instructive to discuss the pattern of the Bethe roots for the winding states.
It turns out that it depends significantly on the sign of the integer ${\tt w}$.
In particular when the twist parameter ${\tt k}$ is positive, i.e., 
$0<{\tt k}<\frac{1}{2}$ the typical  pattern for ${\tt w}=-1$ is shown in fig.\,\ref{fig30}. 
For this state $s(L)=0$ for any $L$ so that
as $L\to \infty$ it flows to the conformal primary 
whose conformal dimensions are as in \eqref{cdeq1},\,\eqref{cdeq2} with $S^z=s=0$. 
The Bethe numbers in this case are given by
\be\label{BN2}
I_j^{({\rm vac})}=-\tfrac{1}{2}\,(M+1)+j-{\tt w}\, , \ \ \ \ j=1,2,\ldots, M
\ee
with ${\tt w}=-1$. This differs from eq.\,\eqref{BN1} by an overall shift by ${\tt w}$. 
To obtain a winding state with non-zero $s$, one should disbalance the
number of roots on the lines $\Im m(\beta)=0$ and $\Im m(\beta)=\frac{\pi}{2}$ while
keeping the Bethe numbers the same as in \eqref{BN2}, similar to what was discussed
for the case with ${\tt w}=0$.
Also recall that the value of $S^z$ is related to the total number of Bethe roots \eqref{Meq1}.
It is important to note that 
together with the formula for the excitation energy \eqref{asdasd},
the quantization condition  \eqref{quantC1} and the product rule \eqref{omegaeq1}
are valid for these states provided $p$ and ${\bar p}$ are taken as in \eqref{cdeq2}.
\bigskip

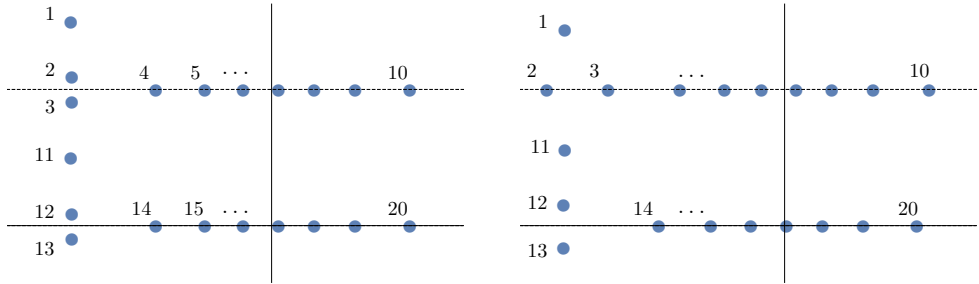
\begin{figure}
\centering
\scalebox{0.75}{
\begin{tikzpicture}
\node at (-4.5,0) {\includegraphics[width=0.53\textwidth]{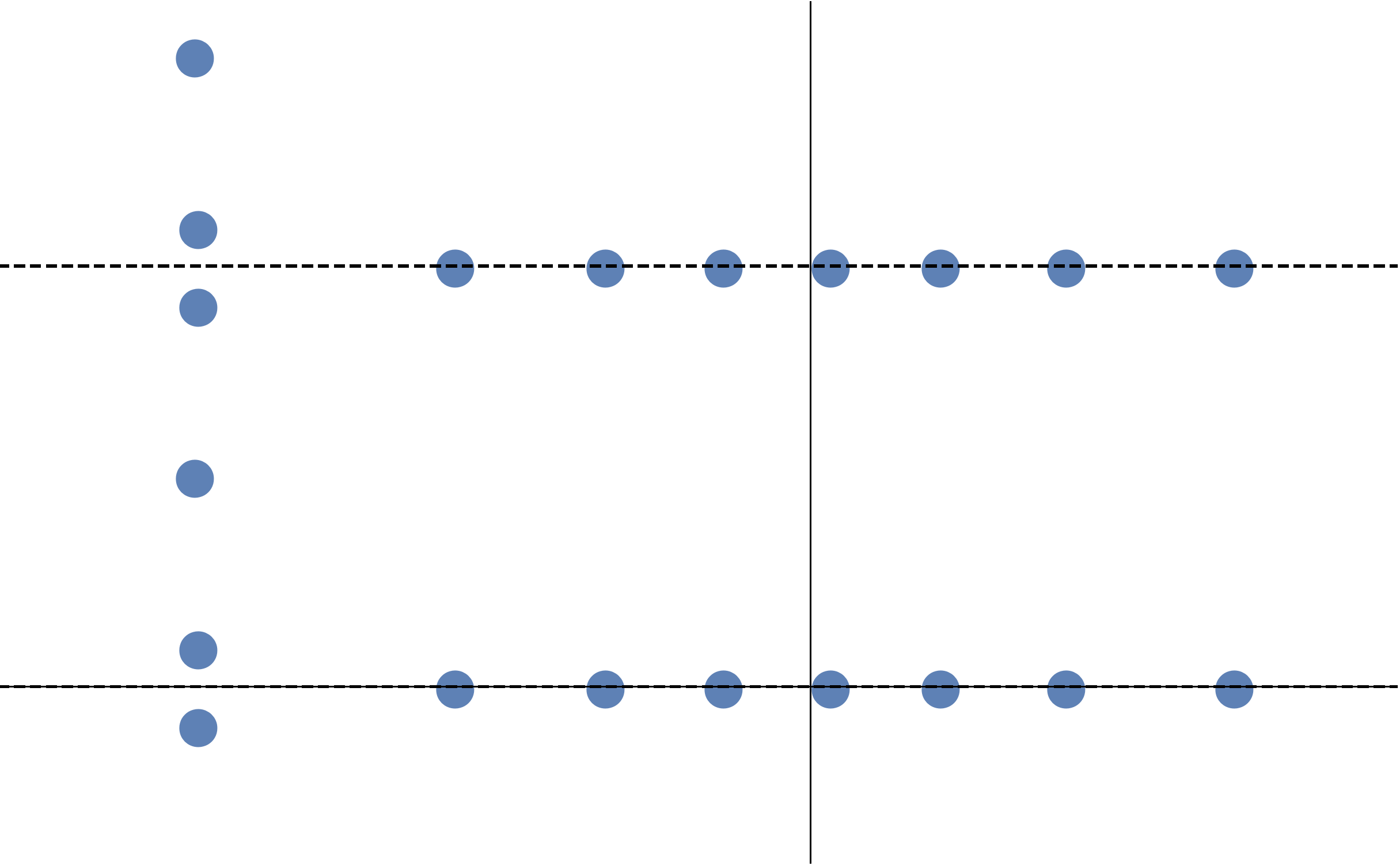}};
\node at (+4.5,0) {\includegraphics[width=0.53\textwidth]{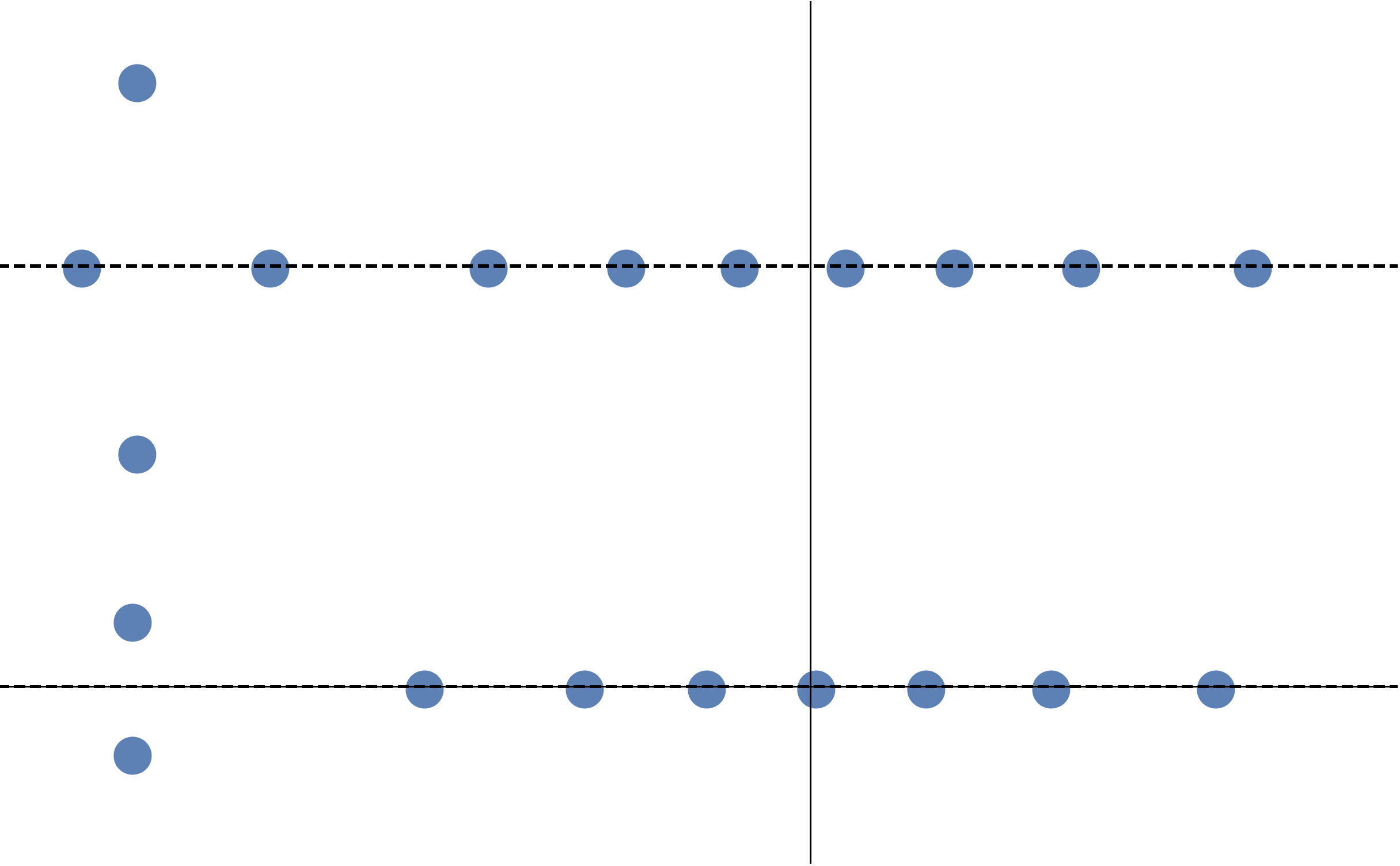}};
\node at (-7.75,2.3) {\small $1$};
\node at (-7.75,1.3) {\small $2$};
\node at (-7.75,0.65) {\small $3$};
\node at (-6.1,1.25) {\small $4$};
\node at (-5.2,1.25) {\small $5$};
\node at (-4.5,1.25) {\large $\ldots$};
\node at (-1.65,1.25) {\small $10$};
\node at (-7.85,-0.2) {\small $11$};
\node at (-7.85,-1.2) {\small $12$};
\node at (-7.85,-1.85) {\small $13$};
\node at (-6.15,-1.15) {\small $14$};
\node at (-5.23,-1.15) {\small $15$};
\node at (-4.5,-1.2) {\large $\ldots$};
\node at (-1.65,-1.15) {\small $20$};
\node at (0.9,2.15) {\small $1$};
\node at (0.7,1.28) {\small $2$};
\node at (1.8,1.28) {\small $3$};
\node at (3.5,1.2) {\large $\ldots$};
\node at (7.5,1.28) {\small $10$};
\node at (0.85,-0.05) {\small $11$};
\node at (0.8,-1.05) {\small $12$};
\node at (0.8,-1.9) {\small $13$};
\node at (2.65,-1.15) {\small $14$};
\node at (3.5,-1.2) {\large $\ldots$};
\node at (7.3,-1.15) {\small $20$};
\end{tikzpicture}
}
\caption{\small
The Bethe roots for the winding states with
${\tt w}=+1$, $m=0$, $S^z=0$ and $L=20$. 
The left panel corresponds
to the state for which $s(L)=0$,
while the state related to the right panel, despite the similar pattern,
has non-zero $s(L)$.
 For both states the  Bethe numbers are as in eqs.\,\eqref{BN2},\,\eqref{BN3b}.
The value of the parameters were taken to be $\gamma=\frac{\pi}{5}$ and ${\tt k}=\frac{1}{25}$. 
\label{fig40}
}
\end{figure}

\begin{figure}[b]
\centering
\scalebox{0.8}{
\begin{tikzpicture}
\node at (-4.5,0) {\includegraphics[width=0.45\textwidth]{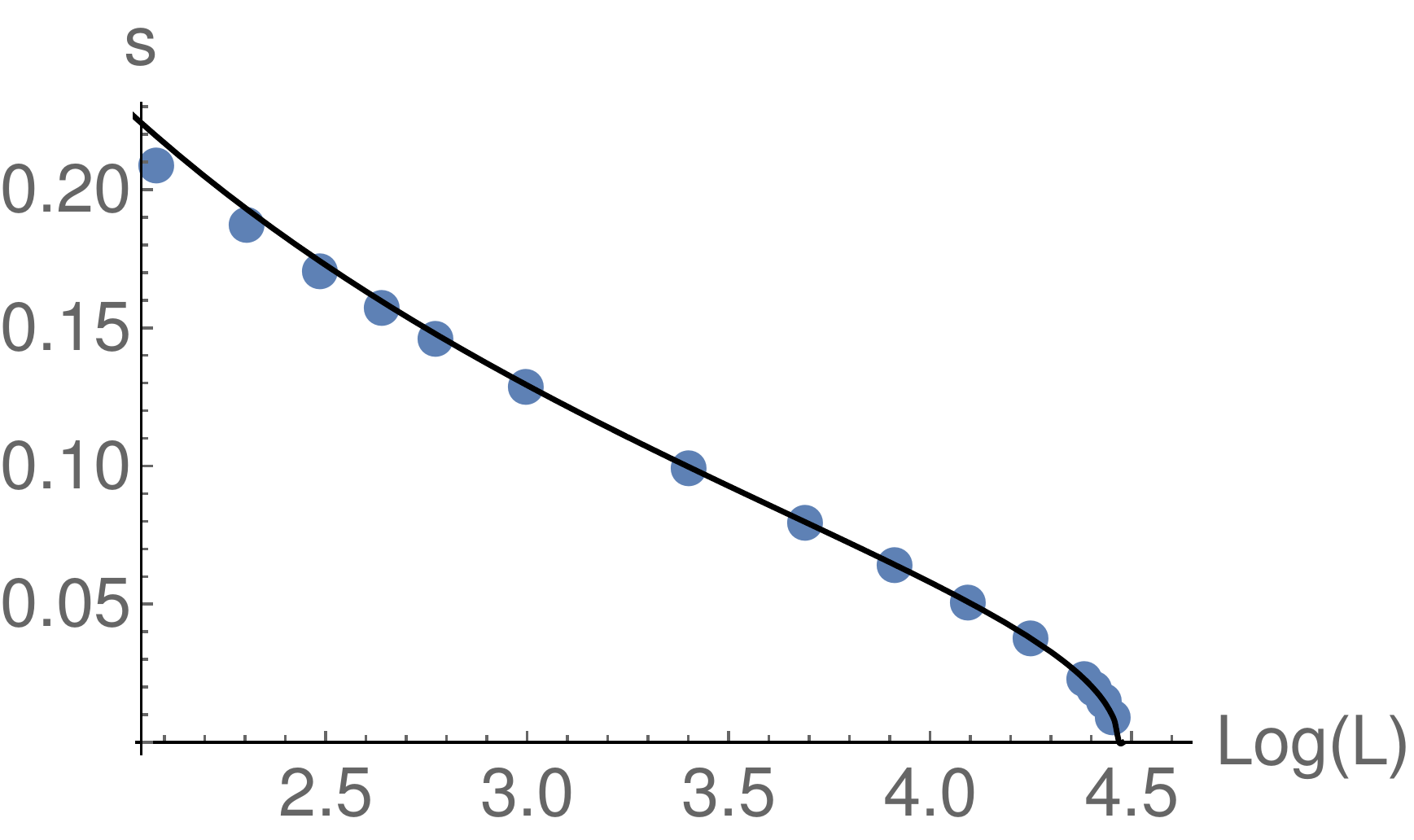}};
\node at (4.5,0.2) {\includegraphics[width=0.45\textwidth]{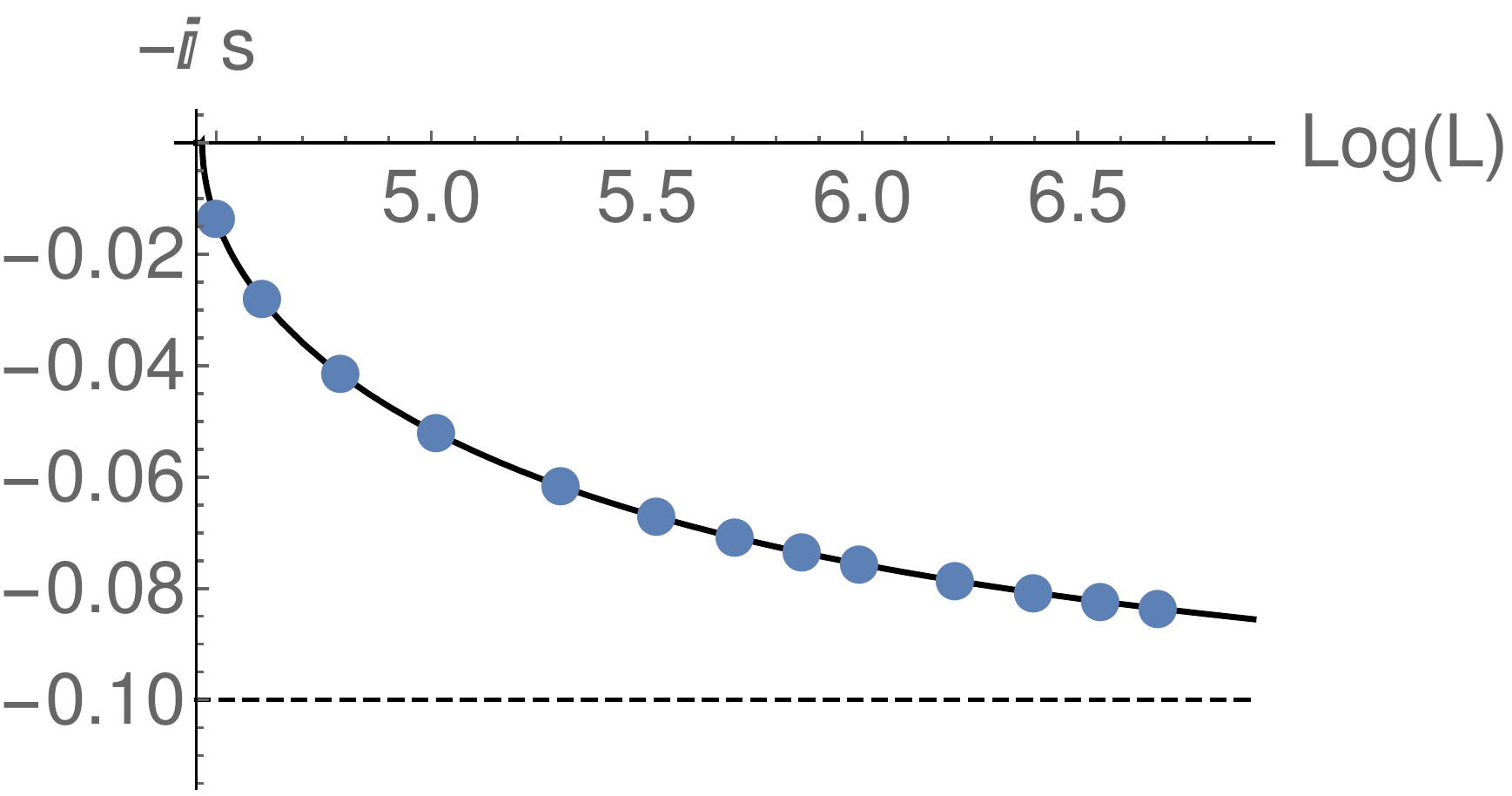}};
\end{tikzpicture}}
\caption{\small
A plot of $s(L)$ coming from the quantization condition \eqref{quantC1} for the case with
${\tt w}=+1$, $m=S^z=0$ with parameters $\gamma=\frac{\pi}{5}$ and ${\tt k}=\frac{1}{25}$. The left panel shows that $s(L)$
vanishes for finite $L$. The right panel continues the flow above the threshold, where $s$ becomes pure
imaginary eventually reaching $s=-\frac{\ri}{10}$ (see eq.\,\eqref{slimeq1}), shown by the dashed line, in the $L\to\infty$ limit.
The data points represent $s$ obtained from the solution to the BA equations associated with the
RG trajectory whose representative state at $L=20$ corresponds to the right panel of fig.\,\ref{fig40}.
}
\label{fig50}
\end{figure}

The states with positive windings ${\tt w}=+1,+2,\ldots\,$($0<{\tt k}<\frac{1}{2}$)
display some interesting phenomena.
The Bethe state corresponding to ${\tt w}=+1$ and with 
$s(L)=0$ ($B(L)=1$)
is depicted on the left panel in fig.\,\ref{fig40}.
Again as $L\to\infty$ it flows towards the conformal primary
having conformal dimensions \eqref{cdeq1},\,\eqref{cdeq2} with $S^z=s=0$ and ${\tt w}=+1$.
For sufficiently large $L$ the Bethe numbers 
are given by $I_j^{({\rm vac})}+\delta I_j$, where the
variations $\delta I_j$ from the ``vacuum'' distribution \eqref{BN2}
are zero except for the following cases:
\be\label{BN3b}
\delta I_1=-1\,, \ \ \ \ \ \ \ \delta I_2=+1\,, \ \ \ \ \ \ \  \delta I_{\frac{1}{2}M+1}=-1\,, \ \ \ \ \ \ \  \delta I_{\frac{1}{2}M+2}=+1
\ee
(recall that $M$ is assumed to be even).
Apart from this, we observed another state having the same Bethe numbers and
a similar pattern of Bethe roots, shown in the right panel of fig.\,\ref{fig40},
but with $s(L)\ne 0$. Remarkably the $L$ dependence of $s(L)$ is still captured by
the quantization condition \eqref{quantC1} with $m=0$.
An important feature of this dependence, which is plotted in fig.\,\ref{fig50}, is that
$s(L)$ vanishes at finite $L$. In fact, in addition to the state corresponding to the pattern
depicted in the right panel of fig.\,\ref{fig40}, there is another state related to it by
the $\mathbb{Z}_2$-transformation ${\cal D}$. The corresponding pattern
is obtained by shifting the roots in accordance with eq.\,\eqref{Deq2}. These states
form a doublet with the same energy but with opposite signs of $s$. 
Thus there are, indeed, three states corresponding to the three real
 solutions of the quantization condition \eqref{quantC1} for $m=0$ and sufficiently small $L$.
A simple examination of \eqref{quantC1}  shows that if $L$ is increased past the point
where the three solutions merge at $s=0$, one solution remains trivial ($s(L)=0$) while the other
two form a pure imaginary conjugated pair. 
This perfectly matches results obtained from the Bethe Ansatz.
In the right panel of fig.\,\ref{fig50} we compare $-\ri s(L)<0$ taken from the BA solutions versus the prediction
coming from the quantization condition. 
Notice that as $L\to\infty$ the value of $s(L)$ approaches a finite pure imaginary number which,
as it follows from  \eqref{quantC1},\,\eqref{phaseeq1} is given by
\be\label{slimeq1}
\lim_{L\to \infty} s(L)=\ri\, \big(\,\bar{p}+\tfrac{1}{2} - \big[ \bar{p}+\tfrac{1}{2}\big]\,\big)\ ,
\ee 
where $[\ldots]$ stands for the integer part of $\bar{p}+\tfrac{1}{2}$.
 Thus we conclude
 that the CFT describing the scaling behaviour of the alternating spin chain
 must contain primary fields of conformal dimensions \eqref{cdeq1} with $s$ not
 only real, but also taking a discreet set of pure imaginary values at the very least.\footnote{%
 A similar phenomenon was observed in regime III of the Izergin-Korepin spin chain \cite{IKspinchain3}.
 We are grateful to Hubert Saleur for drawing our attention to this interesting paper.
 }
 \bigskip

\begin{figure}
\centering
\scalebox{0.7}{
\begin{tikzpicture}
\node at (-5,0) {\includegraphics[width = 0.53\textwidth]{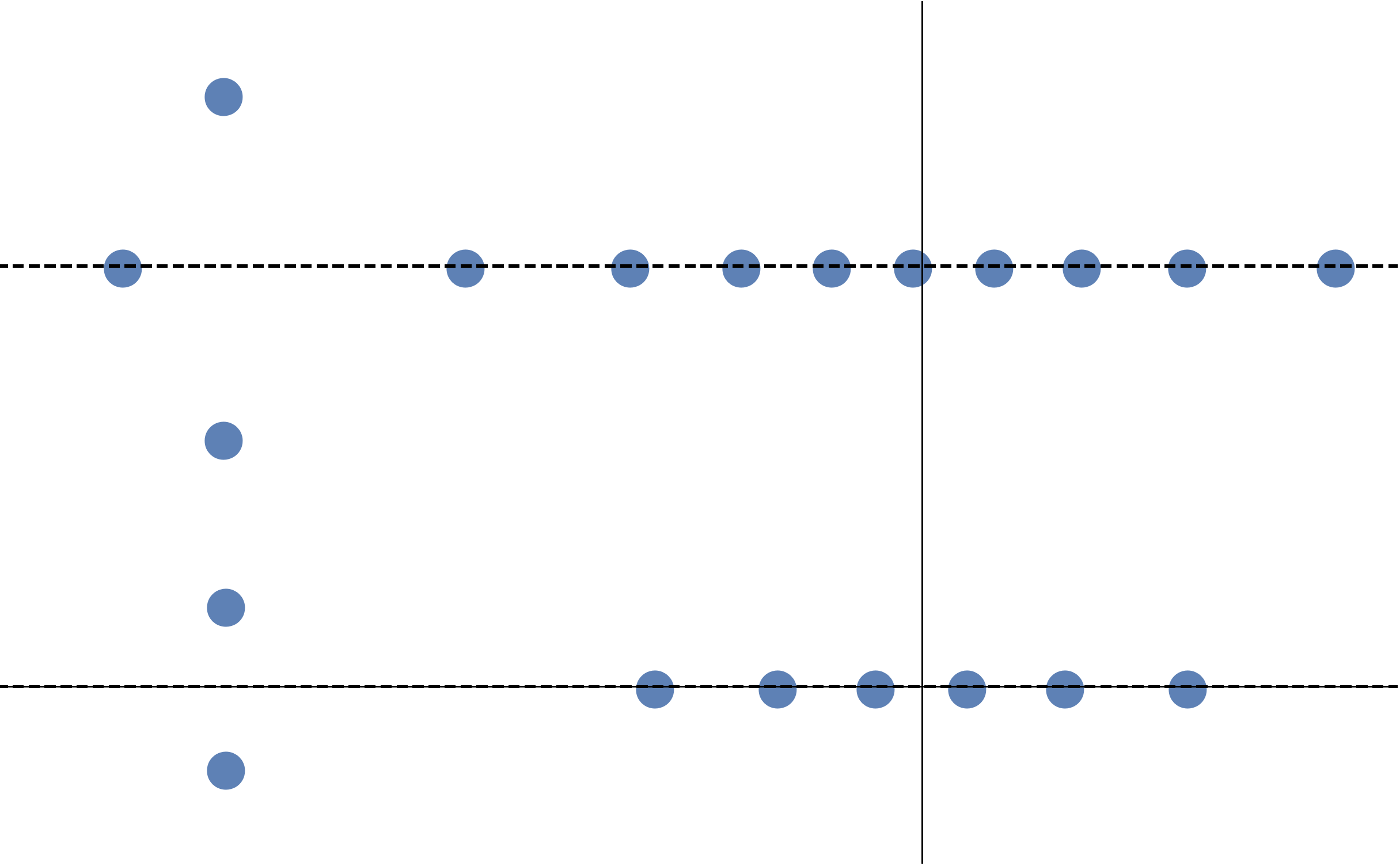}};
\node at (4.5,0) {\includegraphics[width = 0.53\textwidth]{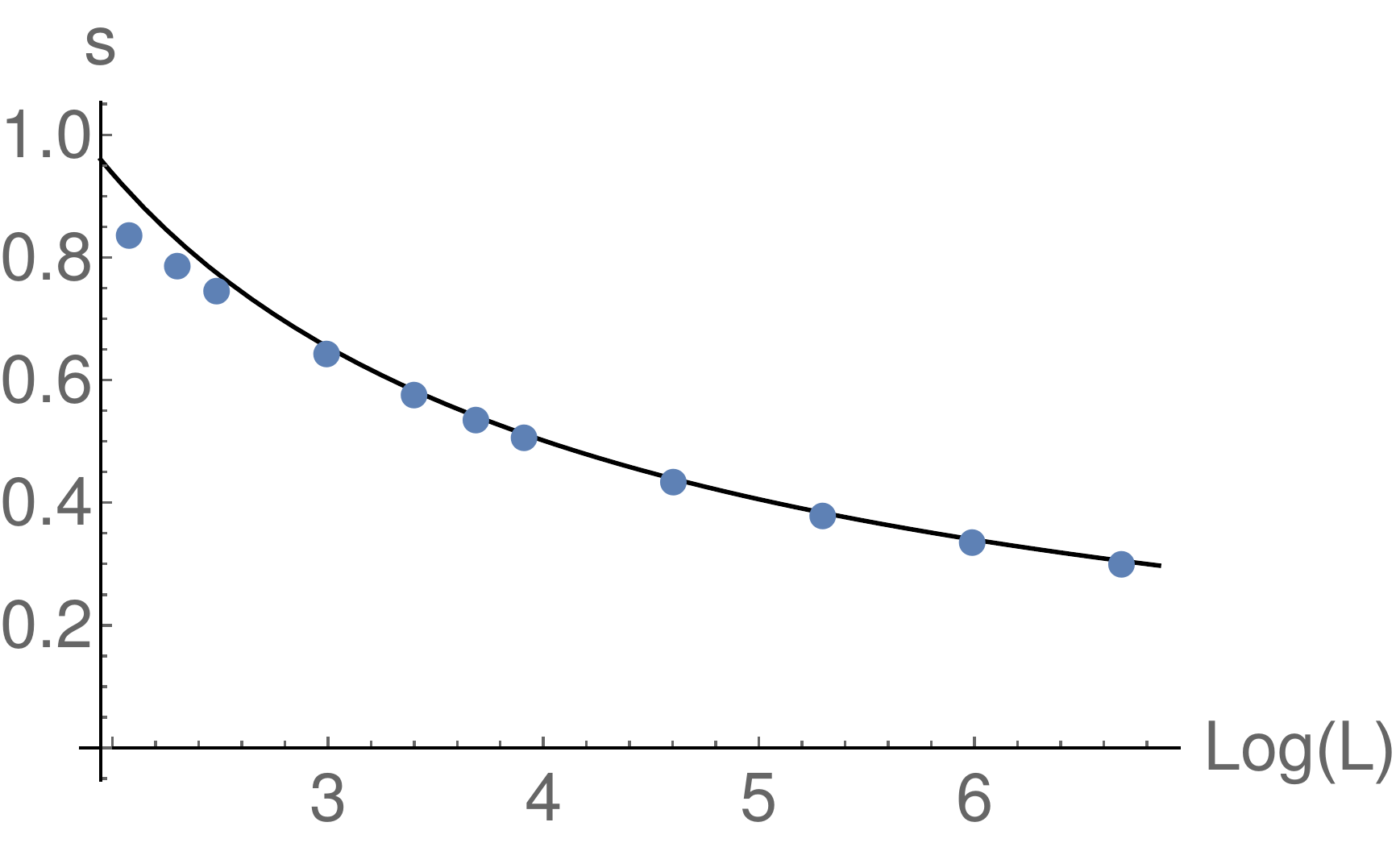}};
\node at (-8,2.1) {\small $1$};
\node at (-8.5,1.2) {\small $2$};
\node at (-6.6,1.2) {\small $3$};
\node at (-4.7,1.2) {\large $\ldots$};
\node at (-1.7,1.2) {\small $11$};
\node at (-8.1,0.15) {\small $12$};
\node at (-8.15,-0.85) {\small $13$};
\node at (-8.15,-1.85) {\small $14$};
\node at (-5.6,-1.2) {\small $15$};
\node at (-4.7,-1.2) {\large $\ldots$};
\node at (-2.5,-1.2) {\small $20$};
\end{tikzpicture}
}
\caption{\small On the left panel,
the pattern of Bethe roots for the winding state with $S^z=0$, ${\tt w}=+1$ and $L=20$.
Data for $s(L)$ computed from the solution to the BA equations for the corresponding 
RG trajectory
is represented by the points in the graph on the right panel of the figure.
The line is a plot of  
$s(L)$ coming from the quantization condition \eqref{quantC1} with $m=2$. The parameters
were taken to be $\gamma=\frac{\pi}{5}$ and ${\tt k}=\frac{1}{25}$.
\label{fig60}
}
\end{figure}
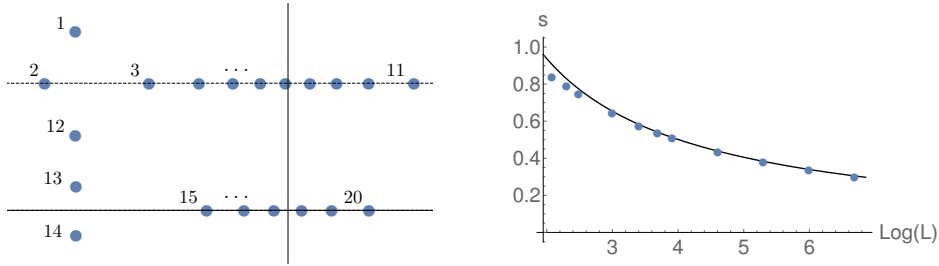

The construction of other winding states with  ${\tt w}=+1$ 
and real non-zero $s(L)$ is achieved by  disbalancing
 the number of roots at $\Im m(\beta)=0$ and $\frac{\pi}{2}$,
 somewhat similar to what was mentioned before for negative ${\tt w}$.
We illustrate the pattern of Bethe roots for such a state
on the left panel of  fig.\,\ref{fig60}. Notice that the flow $s(L)$ agrees with the quantization condition
for $m=2$ (see the right panel of fig.\,\ref{fig60}). 
The Bethe numbers for this state 
turn out to be different from the  case with $m=0$ described by 
eqs.\,\eqref{BN2} and \eqref{BN3b}. Namely, the non-vanishing variations of 
the Bethe numbers from the vacuum distribution \eqref{BN2} are given by
\be
\delta I_1=-1\,, \ \ \ \ \delta I_2=+1\, .
\ee
It should be remembered that the Bethe numbers depend on the choice of
branches for the functions $P(\beta)$ and $\Theta(\beta)$ appearing in the
BA equations \eqref{klkwqnmsd}. In particular, their values are sensitive to local deformations of the cuts 
depicted in fig.\,\ref{branch1}. For example,
for the winding state with ${\tt w}=+1$ having Bethe numbers \eqref{BN2},\,\eqref{BN3b},
it is possible to 
 arrange for all the $\delta I_j$ to be zero by a small deformation of 
 the branch cuts. Moreover a shift of any one of the Bethe roots $\beta_j\to \beta_j+\ri\pi$
 changes the set $\{I_j\}$.
Because of these ambiguities we do not have a clear picture of how to 
assign the Bethe numbers \emph{a priori} to a particular Bethe state. 
Finally we have checked that for all the states discussed up till now, the product rule \eqref{omegaeq1}
is in excellent agreement with numerical data.

\begin{figure}
\centering
\scalebox{0.7}{
\begin{tikzpicture}
\node at (0,0) {\includegraphics[width=0.57\textwidth]{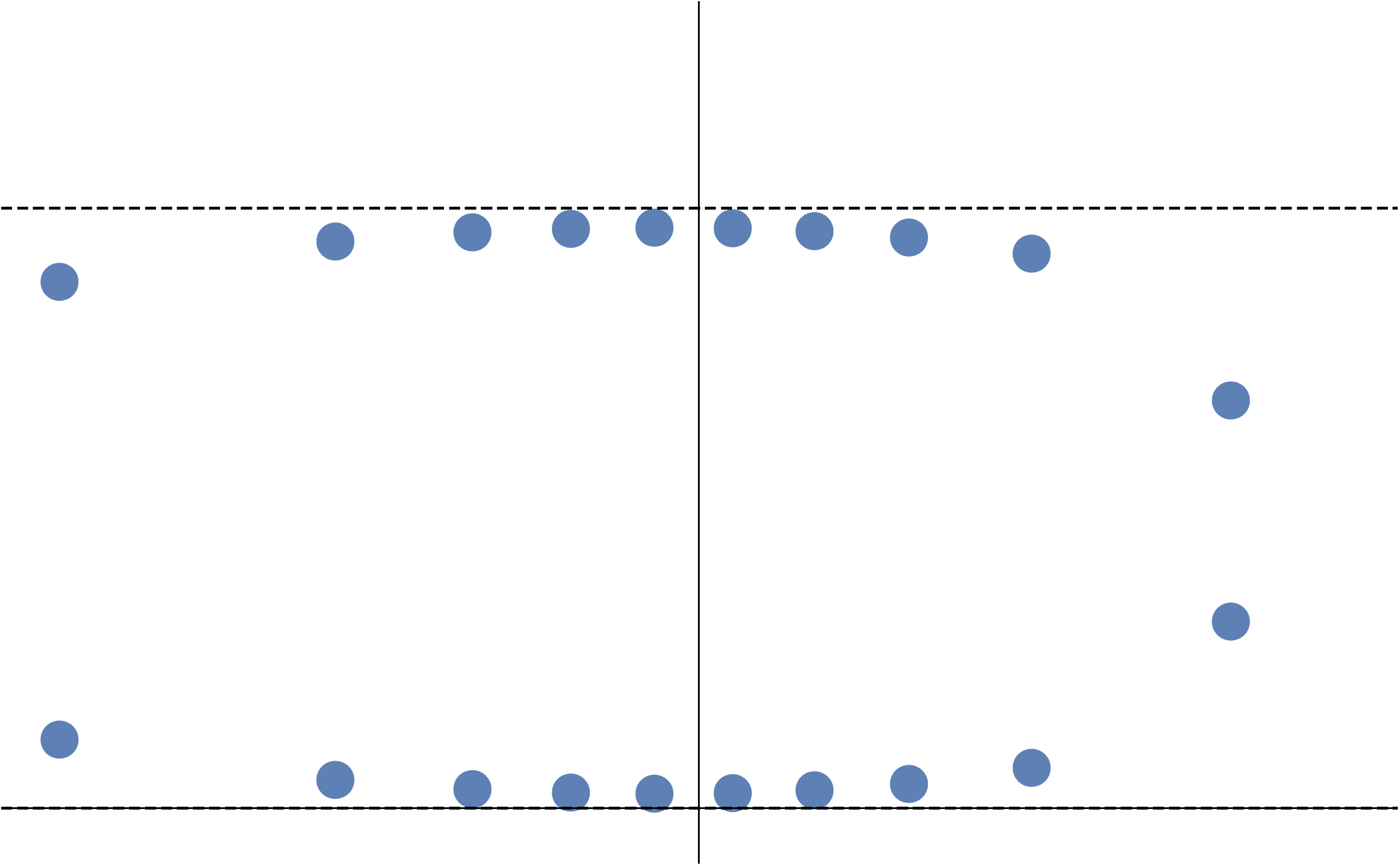}};
\node at (-4.15,1.15) {\small $1$};
\node at (-2.25,1.6) {\small $2$};
\node at (-1.4,1.6) {\small $3$};
\node at (-0.7,1.6) {\large $\ldots$};
\node at (2,1.6) {\small $9$};
\node at (3,0.55) {\small $10$};
\node at (-4.15,-1.55) {\small $11$};
\node at (-2.5,-1.8) {\small $12$};
\node at (-1.55,-1.8) {\small $13$};
\node at (-0.7,-1.8) {\large $\ldots$};
\node at (1.8,-1.75) {\small $19$};
\node at (3,-0.8) {\small $20$};
\end{tikzpicture}
}
\caption{\small The Bethe roots for the state flowing
to a conformal descendent of dimensions $(\bar{\Delta},\Delta+1)$.  
In this case the non-zero variations of the Bethe numbers from the reference distribution
\eqref{BN2} are given by $\delta I_{\frac{M}{2}}=1$ (recall that $M$ is assumed to be even). Note that the set of Bethe roots
 obtained through the ${\cal C}{\cal P}{\cal T}$-transformation of the depicted set has $\delta I_M=1$.
The value of the parameters used for the plot are $\gamma=\frac{\pi}{5}$, $S^z=0$, ${\tt  k}=\frac{1}{20}$
and $L=20$.}
\label{fig70}
\vspace{-0.3cm}
\end{figure}
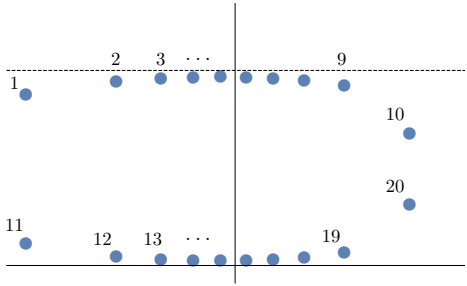
Having described the low energy states of the spin chain that flow to the conformal primaries,
we now turn to the Bethe states that scale to the conformal descendents characterized by
the pair of conformal dimensions $(\bar{\Delta} +{\bar N},\Delta +N)$.
Let's first illustrate them
in the case where the primary conformal dimensions $\bar{\Delta}$, ${\Delta}$ are given by eq.\,\eqref{cdeq1} with $s=0$ and
$p=-\bar{p}=\frac{1}{2}(n+2)\,{\tt k}$, i.e., $S^z={\tt w}=0$. Note that the value of $s(L)$
for the ``descendent'' Bethe states, defined by \eqref{Beq1},\,\eqref{seq1},
turns out to be a complex number in general that becomes zero only in the limit $L\to\infty$.
The pattern of Bethe roots corresponding to the state which flows to the conformal descendent with
 $(\bar{\Delta},\Delta +1)$ is shown in fig.\,\ref{fig70}. Applying the ${\cal C}{\cal P}{\cal T}$-transformation \eqref{CPTeq1} one obtains another state characterized by the same
pair of conformal dimensions.

\begin{figure}
\centering
\scalebox{0.73}{
\begin{tikzpicture}
\node at (-4.5,0) {\includegraphics[width=0.53\textwidth]{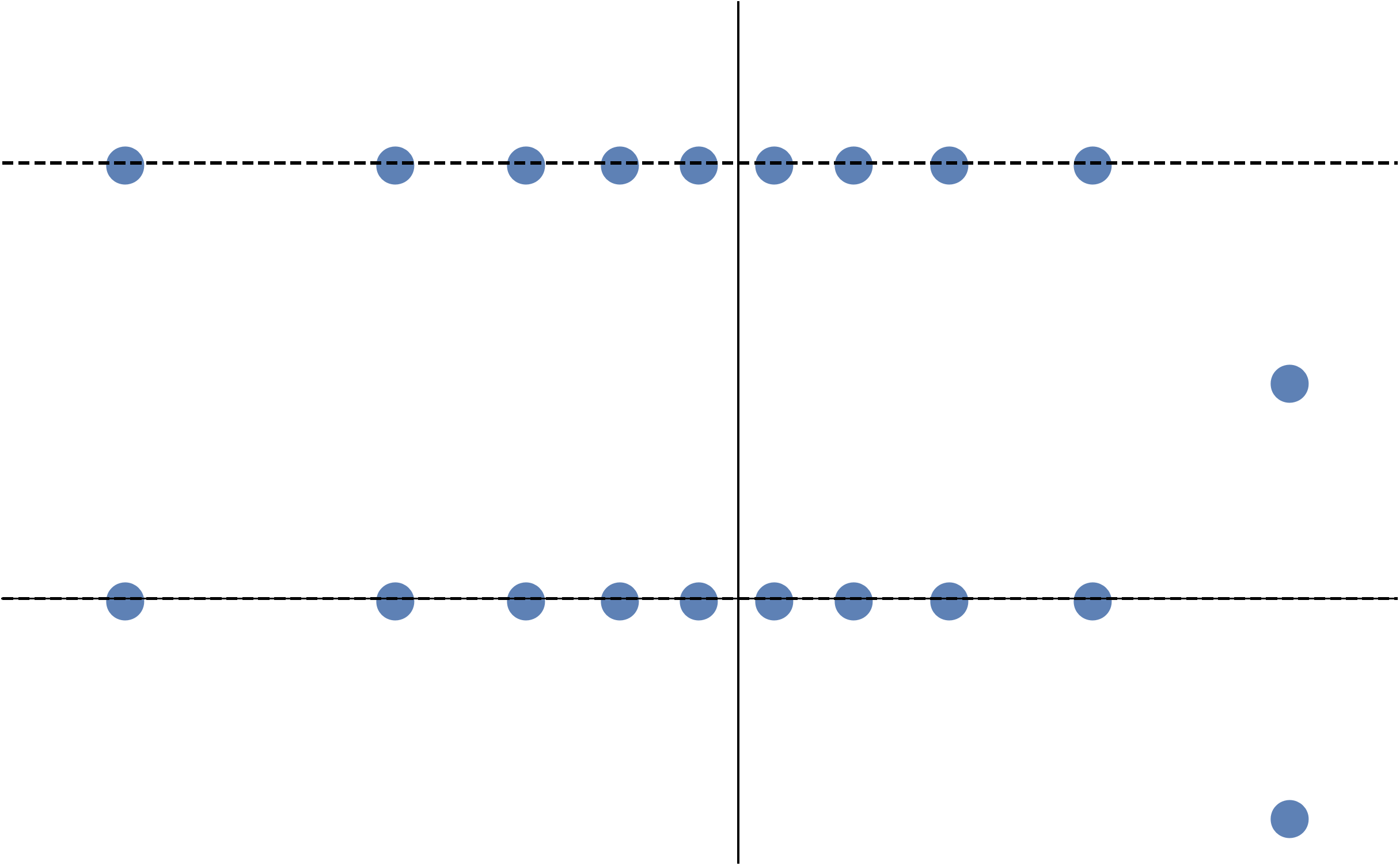}};
\node at (4.5,0.3) {\includegraphics[width=0.53\textwidth]{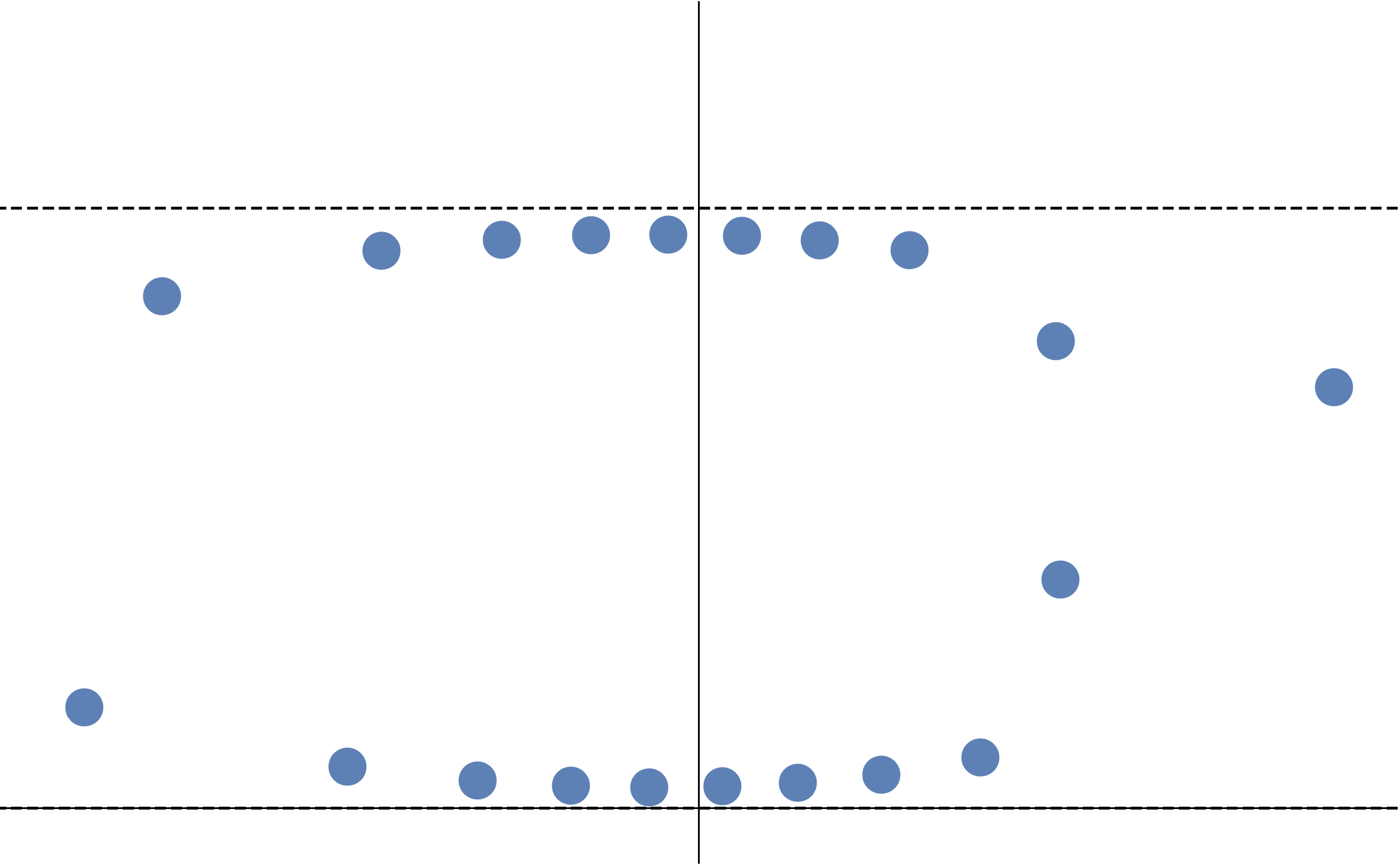}};
\node at (-7.8,1.8) {\small $1$};
\node at (-6.4,1.8) {\small $2$};
\node at (-5.4,1.75) {\large $\ldots$};
\node at (-2.5,1.8) {\small $9$};
\node at (-1.55,0.55) {\small $10$};
\node at (-7.9,-0.6) {\small $11$};
\node at (-6.5,-0.6) {\small $12$};
\node at (-5.5,-0.65) {\large $\ldots$};
\node at (-2.6,-0.6) {\small $19$};
\node at (-1.65,-1.85) {\small $20$};
\node at (1.35,1.3) {\small $1$};
\node at (2.7,1.75) {\small $2$};
\node at (3.4,1.75) {\small $3$};
\node at (4,1.75) {\large $\ldots$};
\node at (6.25,1.1) {\small $9$};
\node at (7.7,0.9) {\small $10$};
\node at (0.8,-0.9) {\small $11$};
\node at (2.3,-1.25) {\small $12$};
\node at (3,-1.3) {\small $13$};
\node at (4,-1.35) {\large $\ldots$};
\node at (5.75,-1.2) {\small $19$};
\node at (6.2,-0.2) {\small $20$};
\end{tikzpicture}
}
\caption{\small The pattern of Bethe roots for the states flowing  to the conformal descendent with
$(\bar{\Delta},\Delta+2)$. The left panel corresponds to the ${\cal C}{\cal P}{\cal T}$-invariant 
state whose non-vanishing $\delta I_j$ are given by $\delta I_{\frac{M}{2}}=\delta I_{M}=1$. For the state
corresponding to the right panel, one has $\delta I_{\frac{M}{2}-1}=\delta I_{\frac{M}{2}}=1$. 
The Bethe roots for the other three states of this type are obtained 
by the  ${\cal C}{\cal P}{\cal T}$ \eqref{CPTeq1} and ${\cal D}$ \eqref{Deq2} transformations.
The non-vanishing
 $\delta I_j$ for these three states are
  $\delta I_{M-1}=\delta I_{M}=1$  (${\cal D}$-transformed state)
 $\delta I_{M}=2$ (${\cal C}{\cal P}{\cal T}$-transformed state) and
  $\delta I_{\frac{M}{2}}=\delta I_{\frac{M}{2}+1}=1$  (${\cal C}{\cal P}{\cal T}$-  and ${\cal D}$-transformed state). 
The values of the parameters used in the plot 
are $\gamma=\frac{\pi}{5}$, $S^z=0$ and ${\tt k}=\frac{1}{30}$.
}
\label{fig80}
\end{figure}
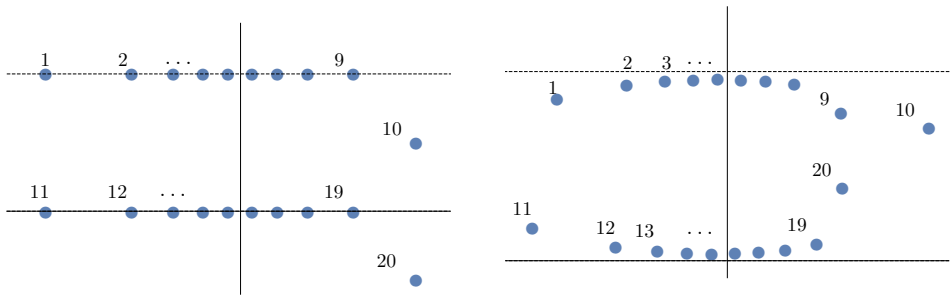

At the second level with $(\bar{\Delta},\Delta +2)$ there exists a ${\cal C}{\cal P}{\cal T}$ -invariant
state for which the typical pattern of Bethe roots is shown in the left panel of fig.\,\ref{fig80}. 
The  roots corresponding to another state at this level are depicted on the right panel of the figure.
Together with its ${\cal D}$-transform, these states form a doublet having the same complex energy.
A doublet with the complex conjugated energy is obtained by means of the ${\cal C}{\cal P}{\cal T}$-transformation.
In the $L\to \infty $ limit, the value of $s$ tends to zero,  the energy becomes real and coincides for all five states.
Of course, a similar picture holds true for the descendent Bethe states associated with
the conformal dimensions $(\bar{\Delta}+1,\Delta)$ and $(\bar{\Delta}+2,\Delta)$.

\bigskip

The major r${\hat{\rm o}}$le in the analysis of the RG flow and 
the identification of the scaling limits of the Bethe states belongs to 
the quantization condition \eqref{quantC1} and the product rule \eqref{omegaeq1}.
The latter is especially important as it typically resolves all the degeneracies
in the spectrum of the Hamiltonian. However there is no reason to expect that  the phase shift $\delta(s)$ appearing in 
the quantization condition eq.\,\eqref{quantC1} is the same for the descendent Bethe states as for those
flowing to the conformal primaries \eqref{phaseeq1}. Of course, $\Omega$ entering the product rule 
\eqref{omegaeq1} must also be modified.
Our approach for determining the asymptotic formula for the product rule and the phase shift for the 
descendent Bethe states was based on the ODE/IQFT correspondence \cite{Dorey:1998pt,Bazhanov:1998wj,Suzuki:2000fc,Bazhanov:2003ni}. 
\medskip

Consider the second order ODE
\bea\label{aisaussfffs}
\bigg[\,-\frac{\rd ^2}{{\rd z}^2}+\frac{p^2-\frac{1}{4}}{z^2}+\frac{2\ri  s}{z}+1
\,
\bigg]\, \Psi=0
\eea
possessing  two singular points -- a regular singular point at $z=0$ and 
an irregular one at $z=\infty$.
The coefficients of the equation are single valued in the vicinity of $z=0$. 
Thus for $p\ne 0,\pm\frac{1}{2}, \pm1,\ldots$, one can introduce a fundamental set of solutions $\Psi_{\pm p}(z)$
such that
\be\label{Psieq1}
\Psi_{\pm p}(z)\to z^{\frac{1}{2}\pm p} \ \ \ \ \ \ {\rm as} \ \ \ z\to 0  \ ,
\ee
and the corresponding monodromy matrix along a small contour about zero is diagonal.
We will assume that $p$ is a real number and 
without loss of generality we take it to be positive. 
For our purposes it is sufficient to focus on the solution $\Psi_{p}(z)$.
In the vicinity of the irregular singular point it has the following behaviour
\bea\label{Psieq2}
\Psi_{p}(z)\to \begin{cases}
C^{(+)}_p\  (+z)^{+\ri s}\ \re^{+z}\ \ \ \ \ &{\rm as}\ \ \ \   \Re e(z)\to+\infty \\[0.2cm]
C^{(-)}_p\  (-z)^{-\ri s}\ \re^{-z}\ \ \ \ \ &{\rm as}\ \ \ \   \Re e(z)\to-\infty
\end{cases}\ .
\eea
Since \eqref{aisaussfffs}  is essentially
the confluent hypergeometric equation,
the formula for the connection coefficients $C^{(\pm)}_p$ is found in any standard textbook.
Using the explicit expression, one observes that the
phase shift $\delta$ \eqref{phaseeq1} and the function $\Omega$ \eqref{omegaeq2}
can be represented in the form
\bea\label{asdasd111}
\delta&=&\frac{16s}{n}\ \log(2)
 -2\ri\log\Bigg[\, \frac{{C}^{(+)}_p\,{C}^{(+)}_{\bar{ p}}}{{C}^{(-)}_p\,{C}^{(-)}_{\bar{p}}}\Bigg]\nonumber \\[-0.2cm]
&& \\[-0.2cm]
\Omega&=&   (n+2)^{\frac{4(\bar{p}-p)}{n+2}}\ 
\Bigg[\frac{\Gamma(1+\frac{2{\bar p}}{n+2})}{\Gamma(1+\frac{2p}{n+2})}\Bigg]^2\ \ 
\frac{ {C}^{(+)}_p\, {C}^{(-)}_p}{{C}^{(+)}_{\bar{p}}\,{C}^{(-)}_{\bar{p}}}\ .\nonumber
\eea
\bigskip

It is well known that the ODE \eqref{aisaussfffs} is obtained from the hypergeometric equation, i.e., 
the Fuchsian differential  equation with three regular singular points,
through a certain limiting procedure. 
The same procedure can be applied to the  so-called generalized hypergeometric oper
introduced and studied in the work \cite{Bazhanov:2013cua}.
This allows one to determine formulae for $\delta$ and $\Omega$ analogous to \eqref{asdasd111} that work
for the descendent Bethe states.
An account of the derivation is given below.

As the first step the ODE  \eqref{aisaussfffs} is replaced by the following 
``generalized confluent hypergeometric equation''
\bea\label{ODE2}
\bigg[-\frac{\rd^2}{\rd z^2}+\frac{p^2-\frac{1}{4}}{z^2}+\frac{2\ri s}{z}+1+\sum_{j=1}^N\bigg(\frac{2}{(z-w_j)^2}+\frac{n_j}{z(z-w_j)}\bigg)\, 
\bigg]\, \Psi=0\ .
\eea
Together with the singularities at $z=0$ and $z=\infty$, 
this equation contains an additional $N$ singular points characterized by the $2N$ complex parameters
$(w_j,n_j)$. We require that the additional singularities are ``apparent'' or monodromy free, i.e.,
any solution of eq.\,\eqref{ODE2} remains single valued in their vicinity.
This requirement leads to a system of algebraic constraints imposed on the set
$(w_j,n_j)$ which read explicitly as 
\bea\label{aosaasi}
c_j\, \big(\, \tfrac{1}{4}\,c_j^2- t_0^{(j)}\big)-t_1^{(j)}=0 \qquad \qquad (j=1,2,\ldots,N)\,,
\eea
where $c_j=-n_j/w_j$ and
\bea
t_0^{(j)}&=&\frac{p^2-\frac{1}{4}}{w_j^2}+\frac{2\ri s}{w_j}+1
-\frac{n_j}{w_j^2}+ \sum_{k\not=j}\bigg(\frac{2}{(w_j-w_k)^2}+\frac{n_k}{w_j(w_j-w_k)}\bigg)\nonumber\\[0.1cm]
t_1^{(j)}&=&-2\ \frac{p^2-\frac{1}{4}}{w_j^3}-\frac{2\ri s}{w_j^2}
+\frac{n_j}{w^3_j}- \sum_{k\not=j}\bigg(\frac{4}{(w_j-w_k)^3}+\frac{n_k\,(2w_j-w_k)}{w_j^2\,(w_j-w_k)^2}\bigg)\ .\nonumber
\eea
Assuming that the positions of the singularities $\{w_j\}$ are given, the above equations 
specify the value of the ``accessory parameters'' $\{n_j\}$. 
A remarkable property of this system is that the
$\{n_j\}$, considered as functions of $\{w_j\}$, satisfy the integrability conditions
\be
w_j\,\frac{\partial n_k}{\partial w_j}=w_k\,\frac{\partial n_j}{\partial w_k}\ .\nonumber
\ee
This means that there exists the (multi-valued) generating function $f^{(N)}(w_1,\ldots,w_N)$ such that
\be\label{nweq1}
n_j=-w_j\,\frac{\partial f^{(N)}}{\partial w_j}\ .
\ee
In the domain
\be
|w_1|<|w_2|<\ldots<|w_N|\ ,\nonumber
\ee
the function $f^{(N)}$ can be defined by supplementing \eqref{nweq1} with the recursion
\bea
\lim_{w_N\to-\infty}f^{(N)}(w_1,\ldots,w_N)=f^{(N-1)}(w_1,\ldots,w_{N-1})\ .
\eea
In particular,
\bea\label{Feq1}
f^{(N)}(w_1,\ldots,w_N)=\int_{{\cal C}(w_N,\infty)}\frac{\rd w}{w}\ n_N(w_1,\ldots,w_{N-1},w)+ f^{(N-1)}(w_1,\ldots,w_{N-1})\ .
\eea
The integration contour here starts at the point $w_N$ and goes to infinity in the complex $w$-plane.
Its precise shape is not important but it must be chosen 
to ensure convergence of the integral, i.e., $n_N$ must vanish along
${\cal C}(w_N,\infty)$:
\be
\lim_{w\to\infty}n_N(w_1,\ldots,w_{N-1},w)\big|_{w\in {\cal C}(w_N,\infty)}=0\ .\nonumber
\ee
Thus the generating function is defined unambiguously up to a single constant $f^{(0)}$ that does not
depend on $\{w_j\}$ and, hence, can be set to zero:
\bea\label{Feq3}
f^{(0)}=0\ .
\eea

The relations \eqref{Feq1},\,\eqref{Feq3} are sufficient for the numerical computation of the generating function $f^{(N)}$,
however we do not have a useful analytic formula except when $N=1$. In this case, as it follows from the algebraic equation \eqref{aosaasi}, 
the pair of complex numbers $(w,n)\equiv (w_1,n_1)$ belong to the cubic
\bea\label{algeq1}
(n+2)\ \big((n+1)^2-4p^2\big)-8\ri s\, (n+1)\, w-4 n\, w^2=0\ .
\eea
Explicit computation of the integral \eqref{Feq1} yields
\bea\label{N1eq1}
f^{(1)}=
p\ \log\big({\check R}^{(1)}\big)+ \ri s\,\log\big(\check{D}^{(1)}\big)
-
n+\frac{1}{2}\ \log\bigg(\,\frac{ w^4}{4p^2}\ \big(1-{\check R}^{(1)}\big)\big(1/{\check R}^{(1)}-1\big)\bigg)\ ,
 \eea
where
\bea\label{N1eq2}
{\check R}^{(1)}&=&
\frac{(n+1+2 p-2 w)(n+1+2 p+2 w)}{(n+1-2 p-2 w)(n+1-2 p+2 w)}\\[0.2cm]
\check{D}^{(1)}&=&
\frac{(1+2 p-2\ri s)(1-2 p-2 \ri s)}{(1+2 p+2\ri s)(1-2p+2\ri s) }\ \  
 \frac{2nw-(n+2)(n-2\ri \, s)}{2n w+(n+2)(n+2\ri \, s)}\ .\nonumber
\eea
Note that in the above formulae $n$ should be understood as a function of $w$ as dictated by the cubic equation \eqref{algeq1}.
Also  the partial derivatives of
$f^{(1)}$ w.r.t. the parameters $p$ and $s$
 can be conveniently expressed in terms of the functions ${\check R}^{(1)}$ and $\check{D}^{(1)}$ respectively:
\bea\label{Deq1a}
\bigg(\frac{\partial f^{(1)}}{\partial p}\bigg)_{w,s}=\,\log\big({\check R}^{(1)}\big)\,, \qquad 
\bigg(\frac{\partial f^{(1)}}{\partial s}\bigg)_{w,p}=
\,\ri  \log\big({\check {D}}^{(1)}\big)\ .
\eea

The subject of our interest are the connection coefficients 
$C^{(\pm, N)}_p$ for the generalized confluent hypergeometric equation \eqref{ODE2}
that are defined in the same way \eqref{Psieq1},\,\eqref{Psieq2} as for the original ODE \eqref{aisaussfffs}.
However now the connection coefficients are functions of the $N$ variables $\{w_j\}$ (and of course $s$ and $p$).
The remarkable fact, which was discussed in the context of the generalized hypergeometric oper in \cite{Bazhanov:2013cua},
is that 
\bea\label{CCeq1a}
&&\frac{{  C}_p^{(+,N)}}{{ {C}}_p^{(-,N)}}=2^{2\ri s}\ \frac{\Gamma(\frac{1}{2}+p-\ri s)}{\Gamma(\frac{1}{2}+p+\ri s)}\  
\exp\bigg(-\ri\, \frac{\partial {  f^{(N)}}}{\partial s}\,\bigg)_{w_j,p}\\ 
&&{ { C}}^{(+,N)}_p\,{ { C}}^{(-,N)}_p=\frac{2^{-2p-1} \ \Gamma^2(1+2 p)}
{\Gamma(\frac{1}{2}+p-\ri s) \Gamma(\frac{1}{2}+p+\ri s)}\ 
\exp\bigg( \frac{\partial {  f^{(N)}}}{\partial p}\,\bigg)_{w_j,s}\ .\nonumber
\eea

Up till now $\{w_j\}$ have been treated as the independent variables.
In fact, for the problem at hand, it is useful to regard $\{n_j\}$ as independent, with $\{w_j\}$
considered as functions of this set defined through the algebraic system \eqref{aosaasi}.
In view of eq.\,\eqref{nweq1} one can perform the Legendre transform of
$f^{(N)}$ w.r.t. the variables $\{\log(w_j)\}$. The result will be denoted by ${g}^{(N)}$:
\be\label{Yfunc1}
{g}^{(N)}(n_1,\ldots,n_N)=\sum_{j=1}^N n_j\,\log(w_j)+f^{(N)}\ .
\ee
Here we emphasize the dependence of $g^{(N)}$ on the set $\{n_j\}$.
It also depends on the parameters $s,p$ and, as follows from the general properties of the Legendre transform,
the following substitution can be made in 
eq.\eqref{CCeq1a}
\bea\label{CCeq1}
\bigg(\frac{\partial {  f^{(N)}}}{\partial s}\,\bigg)_{w_j,p}=
\bigg( \frac{\partial {  {g}^{(N)}}}{\partial s}\bigg)_{n_j,p} 
\ ,\ \ \ \  \ \ \
 \bigg(\frac{\partial {  f^{(N)}}}{\partial p}\,\bigg)_{w_j,s}=
\bigg( \frac{\partial {  {g}^{(N)}}}{\partial p}\bigg)_{n_j,s} \ ,
\eea
provided that the connection coefficients  ${ { C}}^{(\pm,N)}_p$  are understood  as functions of the independent variables $\{n_j\}$.
\bigskip

Finally
we need a clone of the ODE \eqref{ODE2}:
\bea\label{ODE2bar}
\bigg[-\frac{\rd^2}{\rd {\bar z}^2}+\frac{\bar{p}^2-\frac{1}{4}}{\bar{z}^2}+\frac{2\ri s}{\bar{z}}+
1+\sum_{j=1}^{\bar{N}}\bigg(\frac{2}{(\bar{z}-\bar{w}_j)^2}+\frac{\bar{n}_j}{\bar{z}(\bar{z}-\bar{w}_j)}\bigg)\bigg]\, {\Psi}=0\ .
\eea
This equation is only notationally different so that the corresponding connection coefficients are given similarly to \eqref{CCeq1a},\eqref{CCeq1}.
Then, as a result of our analysis we obtained the following expression generalizing \eqref{asdasd111}
\bea\label{asdasd1c}
\delta&=&\frac{16s}{n}\ \log(2)
 -2\ri\log\Bigg[\, \frac{{C}^{(+,N)}_p\ \bar{{C}}^{(+,\bar{N})}_{\bar{ p}}}{{C}^{(-,N)}_p\ \bar{{C}}^{(-,\bar{N})}_{\bar{p}}}\Bigg]\
 \Bigg|_{n_j=\bar{n}_j=n}\nonumber \\[-0.2cm]
&& \\[-0.2cm]
\Omega&=&   (n+2)^{\frac{4(\bar{p}-p)}{n+2}}\ 
\frac{\Gamma^2(1+\frac{2{\bar p}}{n+2})}{\Gamma^2(1+\frac{2p}{n+2})}\ \ 
\frac{ {C}^{(+,N)}_p\  {C}^{(-,N)}_p}{\bar{{C}}^{(+,\bar{N})}_{\bar{p}}\ \bar{{C}}^{(-,\bar{N})}_{\bar{p}}}\ \Bigg|_{n_j=\bar{n}_j=n}\ .\nonumber
\eea
In equation \eqref{asdasd1c} the variables $\{n_j\}$ and $\{\bar{n}_j\}$ are set to the same value $n$. 
From this point on ${C}^{(\pm,N)}_p$, $\bar{{C}}^{(\pm,\bar{N})}_{\bar{p}}$ will always be taken
 to mean the connection coefficients for 
the ODEs \eqref{ODE2},\,\eqref{ODE2bar}
with the parameters $\{n_j\}$, $\{\bar{n}_j\}$ restricted in this way.
Note that in this case the algebraic system \eqref{aosaasi} determining the position of the singularities $\{w_j\}$ simplifies to
\bea\label{sksksk1}
4 n\, w_k^2\!\!&+&\!\!8\ri s\, (n+1)\, w_k-(n+2)\ \big((n+1)^2-4p^2\big)\\[0.2cm]
&+&\!\!
4\ \sum_{j\not=k}^N\frac{w_k\, (\, (n+2)^2\, w_k^2- n(2n+5)\, w_k w_j + n(n+1)\, w_j^2\,)}{(w_k-w_j)^3}=0\  \ \ \ \ \ \ \ \ (k=1,\ldots,N)\ .\nonumber
\eea
A similar equation holds with the set 
$\{w_j\}_{j=1}^N$ replaced by $\{\bar{w}_j\}_{j=1}^{\bar{N}}$ and $p$ substituted with $\bar{p}$.

\bigskip
In view of \eqref{CCeq1a},\,\eqref{CCeq1},
the structure of the formulae \eqref{asdasd1c}  suggests
the following re-writing
\bea\label{hasysay}
\delta&=&-2\, \bigg(\frac{\partial {G}^{(N)}}{\partial s}\bigg)_{n,p}-2\, \bigg(\frac{\partial 
{\bar  G}^{({\bar N})}}{\partial s}\bigg)_{n,{\bar p}}\\[0.3cm]
\Omega&=&\exp\Bigg[\bigg(\frac{\partial {G}^{(N)}}{\partial p}\bigg)_{n,s}-\bigg(\frac{\partial 
{\bar  G}^{({\bar N})}}{\partial {\bar p}}\bigg)_{n,s}\,\Bigg]\ ,\nonumber
\eea
where
\bea\label{iasus1}
G^{(N)}=G^{(0)}(s,p)+g^{(N)}|_{n_j=n}\ ,\ \ \ \  \ \ \ \ 
{\bar  G}^{({\bar N})}=G^{(0)}(s,{\bar p})+{\bar g}^{({\bar N})}|_{\bar{n}_j=n}
\eea
and $G^{(0)}(s,p)$ is   adjusted to make \eqref{hasysay} valid in the case
 $N=\bar{N}=0$ with $g^{(0)}={\bar g}^{(0)}=0$.
The straightforward calculation shows that for $\Re e( p)>-1,\ \Re e( p\pm \ri s)>-\frac{1}{2}$, the function $G^{(0)}$
  can be written as
 \bea\label{iasus2}
{ G}^{(0)}(s,p)&=&
\frac{1}{24}\, \log\Big(4^{n+1}\, n^n\, (n+2)^{-n-2}\Big)
-\frac{2}{n}\ (n+1)\, s^2\ \log(2)
\nonumber \\[0.2cm]
&-&
n\int_0^\infty\frac{\rd t}{t}\ \Bigg[ \, \frac{\re^{-2pt}\sinh(\frac{n t}{2})}{2nt\,\sinh(t)\sinh(\frac{(n+2)t}{2})}+
\frac{\re^{-2pt}\sin^2(st)}{nt\,\sinh(t)}-
\frac{1}{2(n+2) t^2}\\[0.2cm]
&+&\frac{p}{(n+2)\,t}+\bigg(\frac{1}{12}-\frac{p^2}{n+2}-\frac{ s^2}{n}\bigg)\ \re^{-t}\,\Bigg]\ .\nonumber
\eea
The first term here in the r.h.s., independent of $p$ and $s$,  is chosen for future convenience. 
Also note that the integral in \eqref{iasus2} is expressed in terms of the standard 
Barnes' $G$-function.

\bigskip

Let's illustrate the formula \eqref{asdasd1c} for the simplest Bethe states, which flow to the descendent
with conformal dimensions $(\bar{\Delta},\Delta+1)$. In this case the relevant ODEs are
\eqref{ODE2} with one apparent singularity ($N=1$) and the confluent hypergeometric equation, i.e.,
\eqref{ODE2bar} with $\bar{N}=0$. For given values of $n,p$ and $s$ the system \eqref{sksksk1}
boils down to a quadratic for $w\equiv w_1$ whose two solutions are:
\be\label{eqw1}
w_\pm=-\frac{n+1}{2n}\,\Bigg(2\ri\,s\pm\sqrt{n(n+2)}\ \sqrt{1-\frac{4p^2}{(n+1)^2}-\frac{4s^2}{n(n+2)}}\ \Bigg)\ .
\ee
The general relations \eqref{CCeq1a},\,\eqref{asdasd1c} combined with the explicit formulae for $N=1$
\eqref{N1eq1}-\eqref{Deq1a}
 yields
\bea\label{deltaLvl1}
\delta_\pm&=&-2\ri\, \log\bigg[2^{\frac{4\ri s}{n}(n+2)}
\ \frac{\Gamma(\frac{1}{2}+p-{\ri s})\Gamma(\frac{1}{2}+{\bar p}-{\ri s})}{\Gamma(\frac{1}{2}+p+{\ri s})\Gamma(\frac{1}{2}+{\bar p}+{\ri s})} \
\frac{(1+2 p-2\ri s)(1-2 p-2 \ri s)}{(1+2 p+2\ri s)(1-2p+2\ri s) }\nonumber\\[0.2cm]
 &\times&\frac{2n w_\pm-(n+2)(n-2\ri \, s)}{2n w_\pm+(n+2)(n+2\ri \, s)}\,
 \bigg] \,,
 \eea
 while
 \bea\label{OmegaLvl1}
\Omega_\pm&=&2^{2({\bar p}-p)}\ \ 
(n+2)^{\frac{4(\bar{p}-p)}{n+2}} \ \Bigg[\frac{\Gamma\big(1+\frac{2\bar{p}}{n+2}\big)\,\Gamma(1+2p)}
{\Gamma\big(1+\frac{2p}{n+2}\big)\,\Gamma(1+2\bar{p})}\Bigg]^2\ 
\ \frac{\Gamma\big(\frac{1}{2}+\bar{p}+\ri s\big)\,\Gamma\big(\frac{1}{2}+\bar{p}-\ri s\big)}
{\Gamma\big(\frac{1}{2}+p+\ri s\big)\,\Gamma\big(\frac{1}{2}+p-\ri s\big)}\, \nonumber\\[0.2cm]
&\times& \frac{(n+1+2 p-2 w_\pm)(n+1+2 p+2 w_\pm)}{(n+1-2 p-2 w_\pm)(n+1-2 p+2 w_\pm)}\ .
\eea

\begin{figure}
\centering
\scalebox{0.84}{
\begin{tikzpicture}
\node at (-4.6,0) {\includegraphics[width=0.47\textwidth]{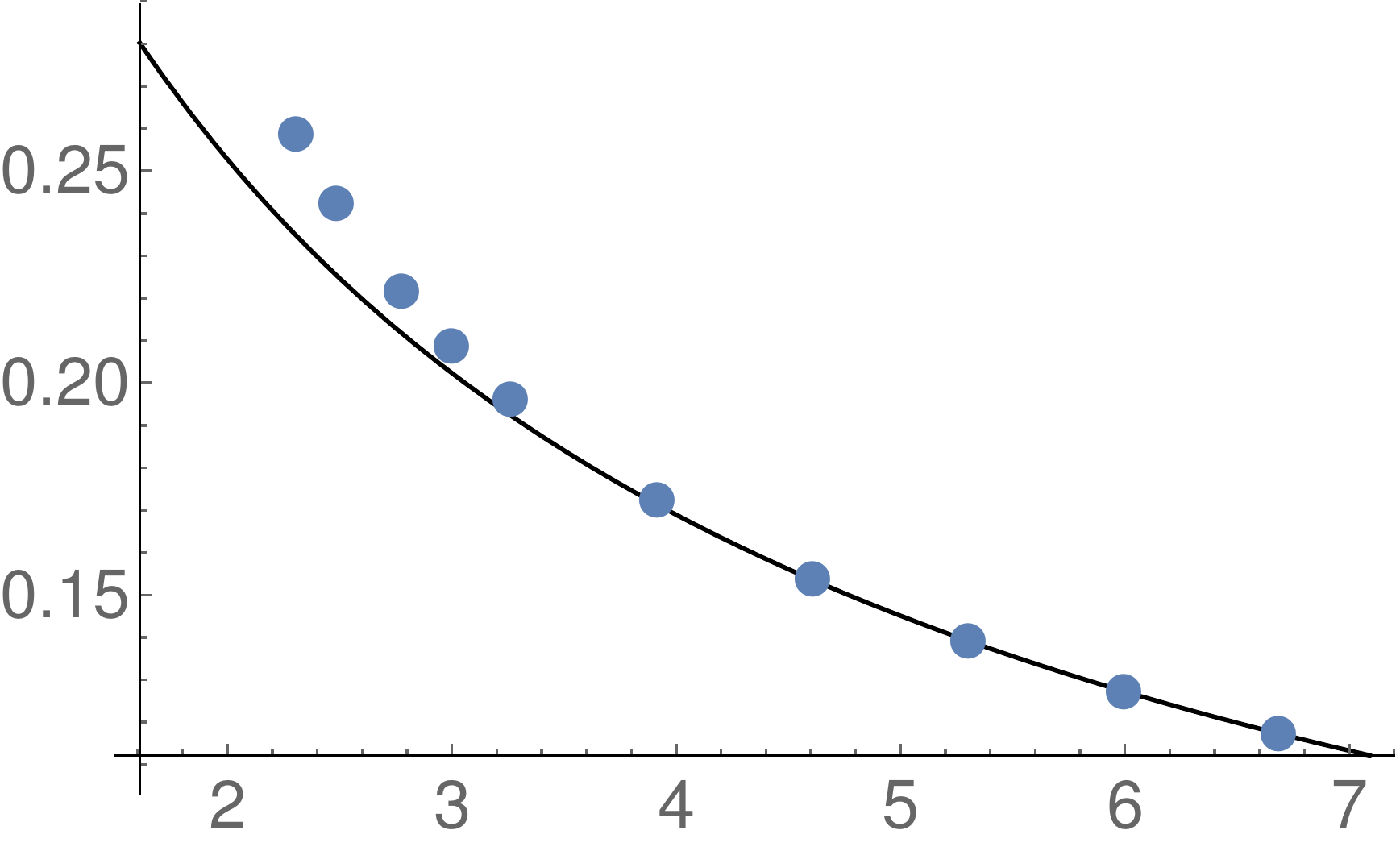}};
\node at (4.6,0) {\includegraphics[width=0.47\textwidth]{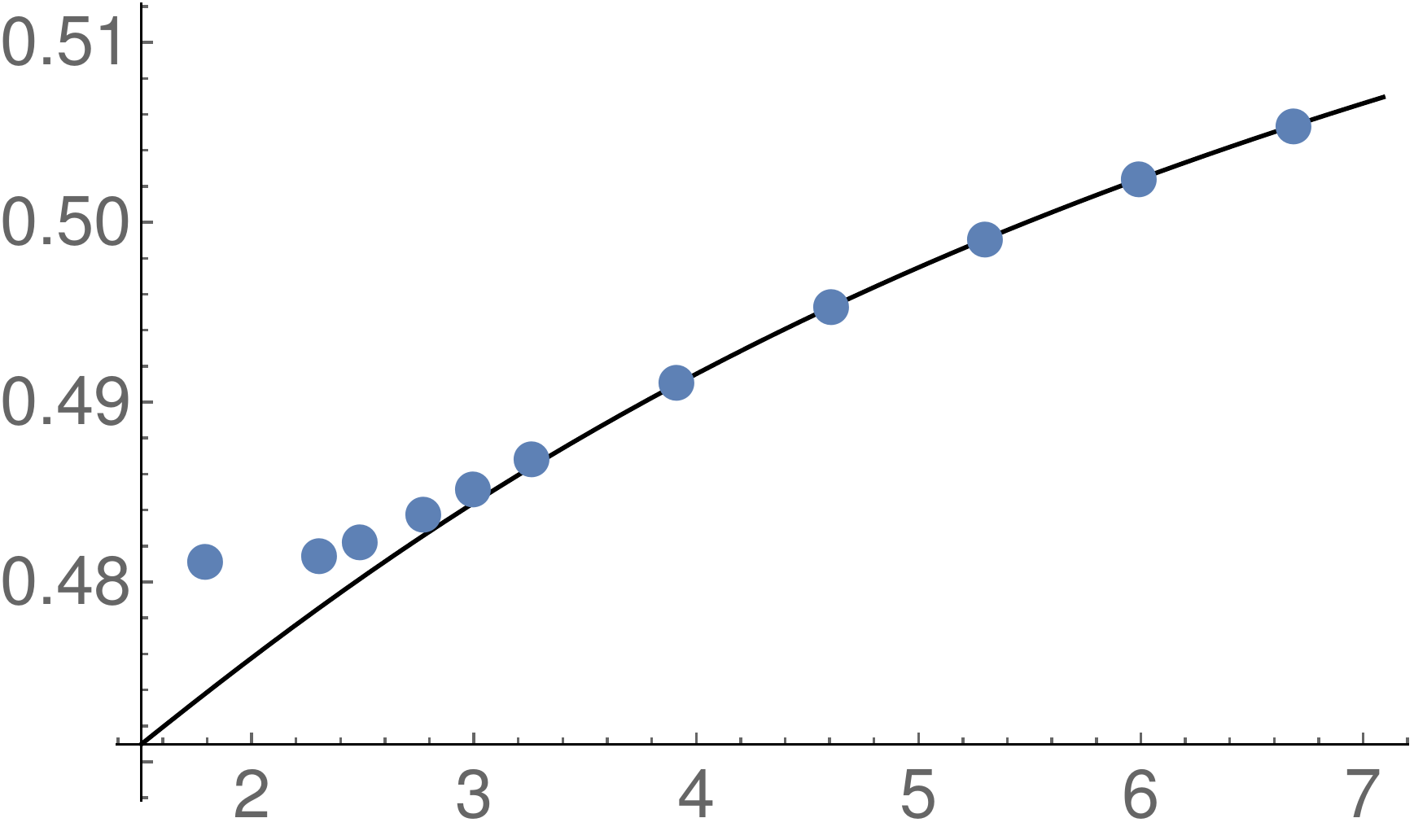}};
\node at (-0.2,-1.7) {$\log(L)$};
\node at (9,-1.7) {$\log(L)$};
\node at (-7.6,2.5) {$-\ri\,s(L)$};
\node at (1.7,2.5) {$L^\nu\, \Pi(L)$};
\end{tikzpicture}
}
\caption{\small
The scale dependence of $s(L)$ (left panel) and $L^{\nu}\,\Pi(L)$
with $\nu=\frac{2n(p-\bar{p})}{n+2}$  (right panel) 
for the state 
flowing to the conformal descendent with dimensions $(\bar{\Delta},\Delta+1)$, whose
typical pattern of Bethe roots is depicted in fig.\,\ref{fig70}.
The points
come from the solution of the BA equations, whereas the solid lines
result from the analytic expressions \eqref{quantC1},\,\eqref{deltaLvl1} for $s(L)$ and 
\eqref{omegaeq1},\,\eqref{OmegaLvl1} for $\Pi(L)$. 
Note that in the right panel, the solid line tends to the finite limit $L^\nu\,\Pi(L)\to 0.554918\ldots $ as $L\to \infty$. 
The integer $m$ 
entering into the quantization condition is chosen to be zero, while the values of the
remaining parameters are $\gamma=\frac{\pi}{5}$, $S^z=0$ and ${\tt k}=\frac{1}{20}$
($p=-\bar{p}=\frac{\tt k}{2}\,(n+2)$, $n=3$).
\label{figa}}
\end{figure}
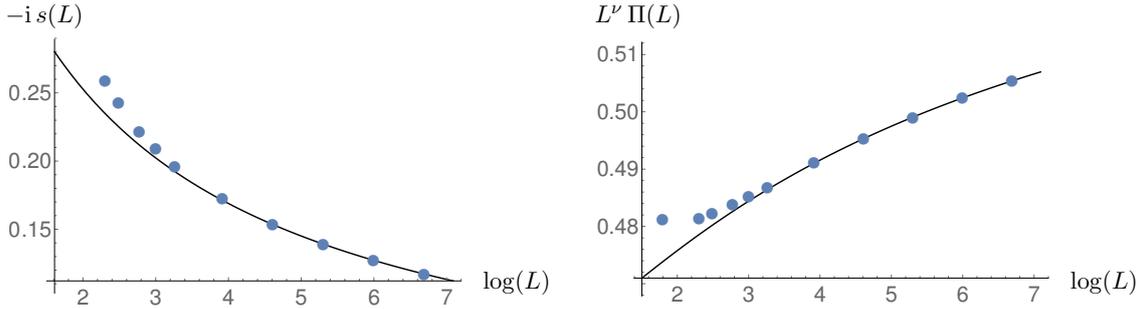

In our discussion of the RG flow we mentioned two states that
become the conformal descendents with $(\bar{\Delta},\Delta+1)$ in the scaling limit.
Fig.\,\ref{fig70} illustrates the pattern of Bethe roots for one such state, while the second
is related to the former by the ${\cal C}{\cal P}{\cal T}$-transformation.
This is in agreement  that for $N=1$ there are two solutions of \eqref{sksksk1}, labeled by
the subscripts ``$\pm$'' in eqs.\,\eqref{eqw1}-\eqref{OmegaLvl1}. It turns out that the
state illustrated in fig.\,\ref{fig70} corresponds to the case of ``+'' with the values of the parameters
taken as in the caption. In the left panel of fig.\,\ref{figa} we compare the numerical data  for $s(L)$
obtained using the BA equations with the predictions coming from the quantization condition 
\eqref{quantC1}, where $m=0$ and the phase shift is taken to be $\delta=\delta_+$ \eqref{deltaLvl1}. 
Numerical results for $\Pi(L)$ \eqref{Pieq1}
are compared with the analytical predictions  \eqref{omegaeq1} with $\Omega=\Omega_+$ \eqref{OmegaLvl1} 
on the right panel of fig.\,\ref{figa}. 
\bigskip

For the case with $N=2$ we found that for general $s,p$ and $n$,
the algebraic equation \eqref{sksksk1}
possesses exactly five solutions up to the permutation $w_1\leftrightarrow w_2$.
We have checked that these solutions correspond to the five Bethe states discussed before that 
 flow to the descendents
with conformal dimensions $(\bar{\Delta},\Delta+2)$.
Unfortunately we were not able to derive explicit analytical expressions for $\Omega$ 
and the phase shift $\delta$ for $N=2$ similar to \eqref{deltaLvl1},\eqref{OmegaLvl1}.
However eqs.\,\eqref{Feq1},\,\eqref{CCeq1a},\,\eqref{asdasd1c} are sufficient 
for their numerical computation. Some results of our analysis are presented in tab.\,\ref{tab10}

\begin{table}

\centering

\begin{tabular}{|c|c|c|c|}
\hline
& & & \\[-0.4cm]
$L$ & $s(L)$ & $\delta(s)$ & {\rm r.h.s.\ of} eq.\,\eqref{quantC1}\\[0.1cm]
\hline
& & & \\[-0.4cm]
$20$ & $-0.042685+0.271838\, \ri $&$1.282157 - 6.469176\,\ri$ & $(1.9 + 4.8\,\ri)\times 10^{-1} $\\[0.1cm]
\hline
& & & \\[-0.4cm]
$50$ &$-0.038249 + 0.221127\,\ri$ & $1.297885 - 7.177411\,\ri$ &$(3.9+9.9\,\ri)\times 10^{-2} $\\[0.1cm]
\hline
& & & \\[-0.4cm]
$100$ & $-0.034314 + 0.196574\,\ri$ &$1.330870 - 7.529820\,\ri$ &$(1.1+2.9\,\ri)\times 10^{-2}$ \\[0.1cm]
\hline
& & & \\[-0.4cm]
$200$ &$-0.030893 + 0.177611\,\ri$ &$1.362463 - 7.806222\,\ri$ & $(3.2+8.2\,\ri)\times 10^{-3}$ \\[0.1cm]
\hline
& & & \\[-0.4cm]
$400$ &$-0.028035 + 0.162200\,\ri $ &$1.389822 - 8.033551\,\ri$ &$(0.9+2.3\,\ri)\times 10^{-3}$ \\[0.1cm]
\hline
& & & \\[-0.4cm]
$800$ &$-0.025647 + 0.149325\,\ri$ &$1.413107 - 8.225372\,\ri$ &$(2.5+6.5\,\ri)\times 10^{-4}$ \\[0.1cm]
\hline
\end{tabular}
\caption{\small The value of $s(L)$ here is listed for the Bethe state, which flows to the conformal descendent
with dimensions $(\bar{\Delta},\Delta+2)$. This state is not the ${\cal C}{\cal P}{\cal T}$-invariant one
and its typical pattern of Bethe roots is shown in the right panel of fig.\,\ref{fig80}.
The phase shift $\delta(s)$ was computed using \eqref{asdasd1c},\,\eqref{CCeq1a} with the function $f^{(2)}$
calculated numerically via eqs.\,\eqref{Feq1},\,\eqref{Feq3}. The last column shows the r.h.s. of the quantization
condition \eqref{quantC1} with $m=0$, where $s$ and $\delta(s)$ are taken from the previous two columns.
The parameters used for the table are $\gamma=\frac{2\pi}{9}$, ${\tt k}=\frac{1}{30}$ and $S^z=0$.
\label{tab10}}
\end{table}
\bigskip

We expect that for general $N$ the number of solutions of the algebraic system \eqref{sksksk1} 
(up to mutual permutations of the set $\{w_j\}$) coincides with the number of partitions of $N$ into integer parts
of two kinds ${\tt p}_2(N)$:
\be\label{aisuasau}
\sum_{N=0}^\infty {\tt p}_2(N)\,q^N=\prod_{k=1}^\infty\,\frac{1}{(1-q^k)^2}=1+2\,q+5\,q^2+10\,q^3+20\,q^4+36\,q^5+\ldots \ .
\ee
Our numerical work suggests 
that for sufficiently large $L$ the states in the low energy part of the spectrum  ($\{|\psi\rangle\}$) can be characterized by  
$p=\tfrac{1}{2}\,\big(S^z+({\tt k}+{\tt w})\,(n+2)\big)$,
$\bar{p}=\tfrac{1}{2}\,\big(S^z-({\tt k}+{\tt w})\,(n+2)\big)$,
$s=s(L)$ as well as an unordered set of solutions to eq.\,\eqref{sksksk1} and one to its barred counterpart.
Furthermore, for large $L$ the Bethe states develop the following factorized structure:
\be\label{state1}
|\psi\rangle\approx\big|\{\bar{w}_j\}_{j=1}^{\bar{N}}
\big\rangle_{s,\bar{p}}\otimes\big|\{{w}_j\}_{j=1}^{{N}}\big\rangle_{s,p}\, \ \ \ \ \ \qquad\qquad  (L\gg 1)\ ,
\ee
where $s=s(L)$ and the $L$ dependence is determined through the quantization condition \eqref{quantC1}
with the phase shift $\delta$ given by eq.\,\eqref{asdasd1c}. 
With this hypothesis one can turn to the scaling limit.
As was explained before, 
to take this limit
the integer $m$ should be assigned an $L$ dependence
such that $s$ is kept fixed as $L\to\infty$.
An immediate question is: what are the range of values of $s$ that appear in the scaling limit?
A simple-minded guess is that all real values are allowed. 
Even if this is true, the admissible set must also contain pure imaginary $s$
as we explained before (see, e.g., fig.\,\ref{fig50}). We are not in a position to answer this question in full
at the moment. 
\bigskip

Despite that the global restrictions on $s$ are unknown, 
assuming that the pair of parameters $(s,p)$ are admissible,
the structure of the linear span
of the chiral states of the form $|\{w_j\}_{j=1}^N\rangle_{s,p}$ seems to be clear.
They form a highest weight representation of the $W_\infty$-algebra, which was studied in \cite{Bakas:1991fs},
with central charge
$
c=\frac{2(n-1)}{n+2}
$.
The latter contains the holomorphic currents $W_j$ with Lorentz spins $j=2,3,\ldots\ $.
All of these currents can be produced from the operator product expansion of
$W_2$ (which is the holomorphic component of the energy momentum tensor) and
the current $W_3$. The highest weight representation is specified by the highest weights
$(\Delta,\varpi)$, where $\Delta$ is the conformal dimension and $\varpi$ is the value of the zero mode of the  $W_3$-current. 
In terms of $s$ and $p$ the 
 parameterization
of $\Delta$ is given by eq.\,\eqref{cdeq1}, while in the proper normalization of the $W_3$-current,
\be
\varpi=s\,\bigg(\frac{3 n p^2 }{3n+4}-\frac{(2n+3) n }{4(3 n+4)}+s^2\bigg)\ .
\ee
The corresponding $W$-module
will be denoted by ${\cal V}_{s,p}$.
All possible states of the form  $\big|\{w_j\}_{j=1}^N\big\rangle_{s,p}$, where $N=0,1,\ldots$ and $\{w_j\}$ 
are an unordered set of solutions of eq.\,\eqref{sksksk1}, are expected to form a basis of 
${\cal V}_{s,p}$. 
\bigskip

Of course there are many ways to introduce a basis for ${\cal V}_{s,p}$. The special property of the states
$\big|\{w_j\}_{j=1}^N\big\rangle_{s,p}$
 is that they are consistent with  a certain
integrable structure of the $W_\infty$-algebra. The latter can be specified as follows;\footnote{This integrable structure was studied
in refs.\cite{Fateev:2005kx} and \cite{Bazhanov:2008yc} in the domain $n<-2$ and $-2<n<0$, respectively.}
using the $W$-currents one can construct a commuting set of local Integrals of Motion (IM)
$\{\mathbb{I}_k\}$ with spin $k=1,2,\ldots$\ :
\be
[\mathbb{I}_k,\mathbb{I}_j]=0\ .\nonumber
\ee
By local we mean that each $\mathbb{I}_k$ is given by an integral over the local density 
built out of the $W$-currents and their derivatives. 
All the local IM are simultaneously diagonalized in the basis $\big|\{w_j\}_{j=1}^N\big\rangle_{s,p}$
, i.e.,
\be\label{eigeq1}
\mathbb{I}_k\,\big|\{w_j\}_{j=1}^N\big\rangle_{s,p}=I_k^{(N)}({\boldsymbol w})\ \big|\{w_j\}_{j=1}^N\big\rangle_{s,p}\ ,
\ee
where the symbol   ${\boldsymbol w}$ is a shortcut notation   for a  set $\{w_j\}$ that satisfies the system \eqref{sksksk1}.
The eigenvalues of the first three IM read explicitly as
\bea\label{eigeq2}
 {{ I}}_1^{(N)}({\boldsymbol w})&=&{ I}^{(0)}_1\big(s, \sqrt{p^2+(n+2)N}\,\big)\\
 {{ I}}_2^{(N)}({\boldsymbol w})&=&{ I}^{(0)}_2\big(s, \sqrt{p^2+(n+2)N}\,\big)+\frac{3 \ri\,n^2}{3n+4}\ 
 \sum_{k=1}^N w_k\nonumber\\
{{ I}}_3^{(N)}({\boldsymbol w}) &=&{ {I}}^{(0)}_3
\big(s, \sqrt{p^2+(n+2)N}\,\big)-\frac{4n^2}{(5n+6)\,(n+2)}\ \bigg( n\,\sum_{k=1}^N w^2_k-
 \ri\,s\, (n+4)\, \sum_{k=1}^N w_k\bigg)\,,\nonumber
 \eea
where
\bea\label{iasussau}
{ {I}}^{(0)}_1(s,p)&=&-\tfrac{n}{12}+\tfrac{n p^2}{n+2}+s^2=n\,\big(\Delta-\tfrac{1}{12}\big)\nonumber\\
{ { I}}^{(0)}_2(s,p)&=&\tfrac{3 n p^2 s}{3n+4}-\tfrac{(2n+3) n s}{4(3 n+4)}+s^3=\varpi
\\
{ {I}}^{(0)}_3(s,p)&=&\tfrac{n^2 p^4}{(5n+6)(n+2)}-
\tfrac{n^2 p^2}{2(5n+6)}-\tfrac{n^2(n-6)(2n+3)}{240 (5n+6)}+\tfrac{6 np^2 s^2}{5 n+6}-
\tfrac{(3n+4) n s^2}{2(5n+6)}+s^4\ .
\nonumber
\eea
\smallskip

A prominent r$\hat{{\rm o}}$le in the integrable structure belongs to the
chiral $Q$-operator, which acts in the highest weight 
module ${\cal V}_{s,p}$.
It is
obtained from the lattice $Q$-operator by taking an appropriately defined scaling limit.
The construction of this chiral operator will be presented elsewhere as it
 requires a detailed description of the $W_\infty$-algebra 
and its representations.
However its eigenvalue, labeled by the set $\{w_j\}$ (see eq.\,\eqref{state1}), can be described without much effort as it
coincides with the connection
coefficients of a certain ODE. The differential equation is given by
\bea\label{aisausau}
\Bigg[\,-\frac{\rd ^2}{{\rd z}^2}+\frac{p^2-\frac{1}{4}}{z^2}+\frac{2\ri  s}{z}+1+
\sum_{j=1}^N\bigg(\frac{2}{(z-w_j)^2}+\frac{n}{z(z-w_j)}\bigg)+\lambda^{-2-n}\ z^n\,
\Bigg]\, \Psi=0\ ,
\eea
where as before the positions of the singularities $\{w_j\}_{j=1}^N$ satisfy the algebraic system \eqref{sksksk1}.
The ODE \eqref{aisausau} contains a term involving the spectral parameter $\lambda$.
Formally setting $\lambda=\infty$ gives back 
eq.\,\eqref{ODE2} with $n_1=\ldots=n_N=n$.
For finite $\lambda$, since $n>0$, the term $\lambda^{-2-n}\ z^n$ is negligible as
$z\to 0$. Thus, similar as in eq.\,\eqref{ODE2}, one can define the solution $\Psi_p$ with $p>0$ through the asymptotic condition
\be
\Psi_{ p}(z)\to z^{\frac{1}{2}+ p} \ \ \ \ \ \ {\rm as} \ \ \ z\to 0  \ .\nonumber
\ee
 For large $z$ the term $\lambda^{-2-n}\ z^n$ becomes dominant
and we introduce another solution, which 
decays along the positive real axis,  through the WKB asymptotic
\bea
\Xi(z)\to  \Big(\frac{z}{\lambda}\Big)^{-\frac{n}{4}}\ 
\exp\bigg(-\frac{2}{n+2}\ \Big(\frac{z}{\lambda}\Big)^{\frac{n}{2}+1}\bigg)\ \ \ \ \ \ {\rm as}\ \ \ \ z\to+\infty\nonumber
\eea
(here we make the technical assumption that $\lambda>0$ and $n>2$). The eigenvalue of the 
CFT $Q$-operator, depending on the spectral parameter $\lambda$, essentially coincides with the Wronskian
$ W[\Psi_{ p},\Xi]=\Xi\,\partial_z\Psi_{p}-\Psi_{p}\,\partial_z\Xi$
 of 
these two solutions. 
It is not difficult to show that 
\be
\lim_{\lambda\to 0}\,\lambda^{\frac{1}{2}- p}\  { W}[\Psi_{ p},\Xi]= \sqrt{\frac{n+2}{\pi }}\ 
(n+2)^{\frac{2p}{n+2}}\,\Gamma(1+\tfrac{2p}{n+2})\ .\nonumber
\ee
The last formula implies that  $A_p(\lambda)$ defined as
\be
A_p(\lambda)=\sqrt{\frac{\pi}{n+2}}\ 
(n+2)^{-\frac{2p}{n+2}}\ \ \frac{\lambda^{\frac{1}{2}- p}\  W[\Psi_p,\Xi]}{\Gamma(1+\frac{2p}{n+2})}
\ee
satisfies the normalization condition
\be
A_p(\lambda)\to 1\ \ \ \ \ \ \ {\rm for }\ \ \ \ \ \ \ \ \ \lambda\to  0\ .\nonumber
\ee
It turns out that the connection coefficient $A_p(\lambda)$, up to a simple factor, coincides with the eigenvalue of the 
chiral counterpart of the lattice $Q$-operator.
Up to the simple modification
\bea\label{CpmNeq1}
{\mathfrak C}^{(\pm,N)}_p=\sqrt{\frac{\pi}{n+2}}\ (n+2)^{-\frac{2p}{n+2}}\ \ 2^{1\pm\frac{2\ri s}{n}}\ \ 
\frac{C_p^{(\pm,N)}}{ \Gamma(1+\frac{2p}{n+2})}\ ,
\eea
the previously defined connection coefficients $C^{(\pm,N)}_p$ \eqref{CCeq1a} along
  with the eigenvalues of the local IM \eqref{eigeq1}-\eqref{iasussau}
appear in the large-$\lambda$ asymptotic of $A_p(\lambda)$: 
\bea\label{Apeq1}
 A_p(\lambda)&\asymp& {\mathfrak C}^{(+,N)}_p\ \lambda^{+\frac{\ri s(n+2)}{n}-p}\ \exp\Big( b_{-1}\, 
\lambda^{\frac{n+2}{n}}\Big) 
\\ 
&\times&\exp\bigg(\, \sum_{m=1}^\infty(-\ri)^{m+1}\  b_m\ { {  I}}^{(N)}_m\ \lambda^{-\frac{(n+2)m}{n}}
+
\sum_{m=1}^\infty\,  {\tilde { H}}^{(+,N)}_m\
 \lambda^{- (n+2)m} \, \bigg)\nonumber
 \eea
 for $\Re e(\lambda)\to+\infty$ and
 \bea\label{Apeq2}
 A_p(\lambda)&\asymp& {\mathfrak C}^{(-,N)}_p\ (-\lambda)^{-\frac{\ri s(n+2)}{n}-p}\ \exp\Big( b_{-1}\, 
(-\lambda)^{\frac{n+2}{n}}\Big) \\ 
&\times& \exp\bigg(\,
\sum_{m=1}^\infty (+\ri )^{m+1}\,b_m\ { {  I}}^{(N)}_m\ (-\lambda)^{-\frac{(n+2)m}{n}}
+
\sum_{m=1}^\infty\, {\tilde { H}}^{(-,N)}_m\
(- \lambda)^{- (n+2)m} \, \bigg)\nonumber
 \eea
  for $\Re e(\lambda)\to-\infty$.
Here we use  the constants
\bea
b_m&=&
2^{m+1}\ \
\frac{\Gamma(\frac{1}{2}+\frac{n+1}{n}\, m)\Gamma(1-\frac{m}{n})}{2\sqrt{\pi}\, m\, (m+1)!}\ .\nonumber
\eea
The formulae \eqref{Apeq1},\,\eqref{Apeq2} also contain the notation ${\tilde {  H}}^{(\pm,N)}_m$,
which stands for the eigenvalues
of the so-called dual non-local IM. 
 The latter are important elements of the integrable structure, 
 but they will not be touched upon in this paper. 
Note that  ${\mathfrak C}^{(\pm,N)}_p$ \eqref{CpmNeq1}
are also the eigenvalues of certain non-local IM. In particular
their product,
up to an overall normalization depending on $s$ and $p$, coincide with
the eigenvalue of the so-called reflection operator. 
The latter can be defined similarly as in \cite{Bazhanov:2013cua} and of course commutes with the 
local IM:
\be
[{\mathbb R},\mathbb{I}_k]=0\ .\nonumber
\ee
The eigenvalue of the reflection operator,
\be
{\mathbb R}\,\big|\{w_j\}_{j=1}^N\big\rangle_{s,p}=
R^{(0)}\ 
{\check R}^{(N)}({\boldsymbol  w})\ \big|\{w_j\}_{j=1}^N\big\rangle_{s,p}\,,
\ee
is given by
\be
{\check R}^{(N)}({\boldsymbol  w})
= \frac{ {C}^{(+,N)}_p\, {C}^{(-,N)}_p}{ {C}^{(+,0)}_p\, {C}^{(-,0)}_p}
\ee
and
\bea
R^{(0)}=
2^{-4p}\ (n+2)^{-\frac{8 p}{n+2}}\ \ \Bigg[\frac{\Gamma(1-\frac{2p}{n+2}) \,\Gamma(1+2 p)}
{\Gamma(1+\frac{2p}{n+2})\, \Gamma(1-2p)}\Bigg]^2\ \ 
\frac{\Gamma(\tfrac{1}{2}-p+\ri s)\ \Gamma(\tfrac{1}{2}-p-\ri s)}{ \Gamma(\tfrac{1}{2}+p+\ri s)\ \Gamma(\tfrac{1}{2}+p-\ri s)}\ .
\eea

\bigskip

Along with 
the reflection operator  one can consider  the operator $\check{{\mathbb D}}$ whose eigenvalues,
 \be
\check{{\mathbb D}}\,\big|\{w_j\}_{j=1}^N\big\rangle_{s,p}= \check{D}^{(N)}({\boldsymbol  w})\ \big|\{w_j\}_{j=1}^N\big\rangle_{s,p}\,,
\ee
are given by
\be
\check{D}^{(N)}({\boldsymbol  w})=  
\frac{{C}^{(+,N)}_p}{ {C}^{(-,N)}_p}\ \frac{{C}^{(-,0)}_p}{ {C}^{(+,0)}_p}
\ .
\ee
It plays a special r${\hat {\rm o}}$le in calculation of the density of states in the continuous theory. Indeed, assuming that the 
integer $m$ is even,  the quantization condition\ \eqref{quantC1} supplemented by \eqref{asdasd1c}, implies
that the density of the  descendent states with the conformal dimensions $({\bar \Delta}+{\bar N},\Delta+N)$ is related to the density of the
conformal primaries, $\rho_0(s)$,  as follows
\bea
\!\!\!\rho^{(N,{\bar N)}}(s)={\tt p}_2(N)\,{\tt p}_2({\bar N})\  \rho_0(s)+\frac{1 }{2\pi\ri }\  \partial_s\Big( \,
{\tt p}_2({\bar N})\, \log {\rm det}_N(\check{\mathbb D})+{\tt p}_2( N)\, 
 \log {\rm det}_{\bar N}({\bar {\check{\mathbb D}}})\,\Big) .
\eea
Here  ${\tt p}_2(N)$ is the number of partitions as in eq.\eqref{aisuasau} and 
the symbol $\det_N({\check{\mathbb D}})$  stands for  the determinant of the operator  ${\check{\mathbb D}}$ restricted to the  
subspace of  chiral states with conformal dimension $\Delta+N$, i.e.,
\bea
{\rm det}_N(\check{\mathbb D})=\prod_{\boldsymbol w}{\check D}^{(N)}({\boldsymbol  w})\nonumber
\eea
and similar for the barred counterpart.
 The density of the
conformal primaries reads explicitly as
\bea\label{ausasua}
\rho_0(s)=\frac{2}{\pi}\ \Bigg( \!\log(L/L_0)+\frac{1}{4\ri }\ \partial_s \log\bigg[
\ \frac{\Gamma(\frac{1}{2}+p-{\ri s})\Gamma(\frac{1}{2}+{\bar p}-{\ri s})}{\Gamma(\frac{1}{2}+p+{\ri s})\Gamma(\frac{1}{2}+{\bar p}+{\ri s})}\, \bigg]\,\Bigg)\ ,
\eea
where $1/L_0=\frac{ 4^{\frac{n+1}{n}}\Gamma(\frac{3}{2}+\frac{1}{n})}{\sqrt{\pi}\,\Gamma(1+\frac{1}{n})}$.
 It should be pointed out that  eq.\eqref{ausasua} was originally proposed in ref.\cite{Ikhlef:2011ay}. 
 As it follows from  the formula \eqref{N1eq2}, in the case $N=1$ one has
\bea
\bigskip
{\rm det}_1({\check{\mathbb D}})=\frac{(1+2 p-2\ri s)(1-2 p-2 \ri s)}{(1+2 p+2\ri s)(1-2p+2\ri s) }\ .
\eea
In principle it is possible to calculate ${\rm det}_N({\check {\mathbb D}})$ for higher levels.
For example, for $N=2$ we found
\bea
{\rm det}_2({\check{\mathbb D}})=\frac{(1+2p-2\ri s)^2\,(1-2p-2\ri s)^2\,(3+2p-2\ri s)\,(3-2p-2\ri s)}{
                                                     (1+2p+2\ri s)^2\,(1-2p+2\ri s)^2\,(3+2p+2\ri s)\,(3-2p+2\ri s)}\ .
\eea
This illustrates the fact 
that  the density of the conformal descendent states with the conformal dimensions $({\bar \Delta}+{\bar N},\Delta+N)$ 
does not coincide with 
${\tt p}_2(N)\,{\tt p}_2({\bar N})\,\rho_0(s)$.

\begin{figure}
\centering
\scalebox{0.85}{
\begin{tikzpicture}
\node at (-7,2.5) {\small $\frac{L^2}{2\pi v_{\rm F}}\,\delta E_{{\rm odd}}$};
\node at (1.8,2.6) {\small $\frac{L}{2\pi v_{\rm F}}\,\delta E_{{\rm even}}$};
\node at (8.75,-1.6) {\small $\log(L)$};
\node at (-0.6,-1.55) {\small $\log(L)$};
\node at (-4.8,0) {\includegraphics[width=0.45\textwidth]{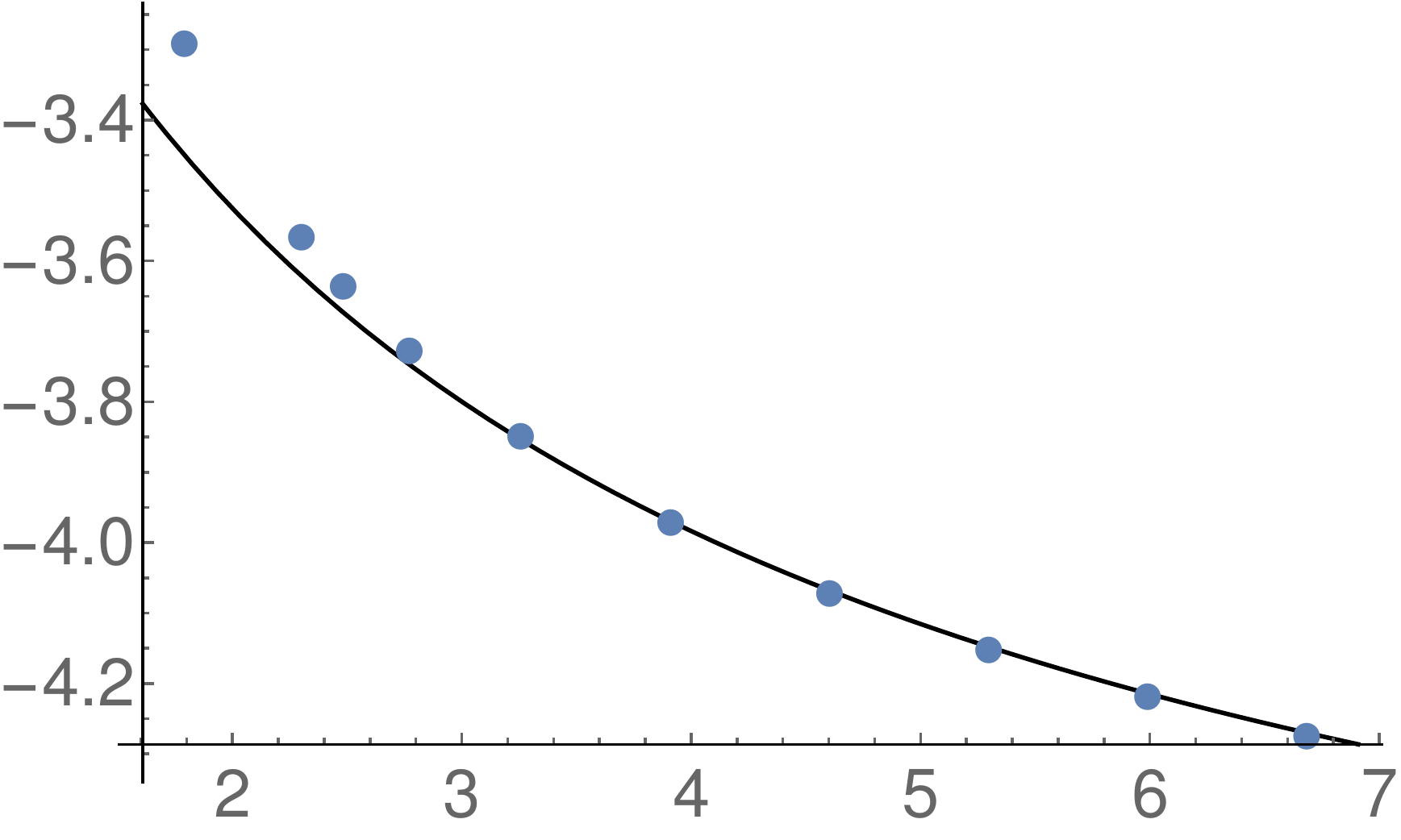}};
\node at (+4.8,0.05) {\includegraphics[width=0.43\textwidth]{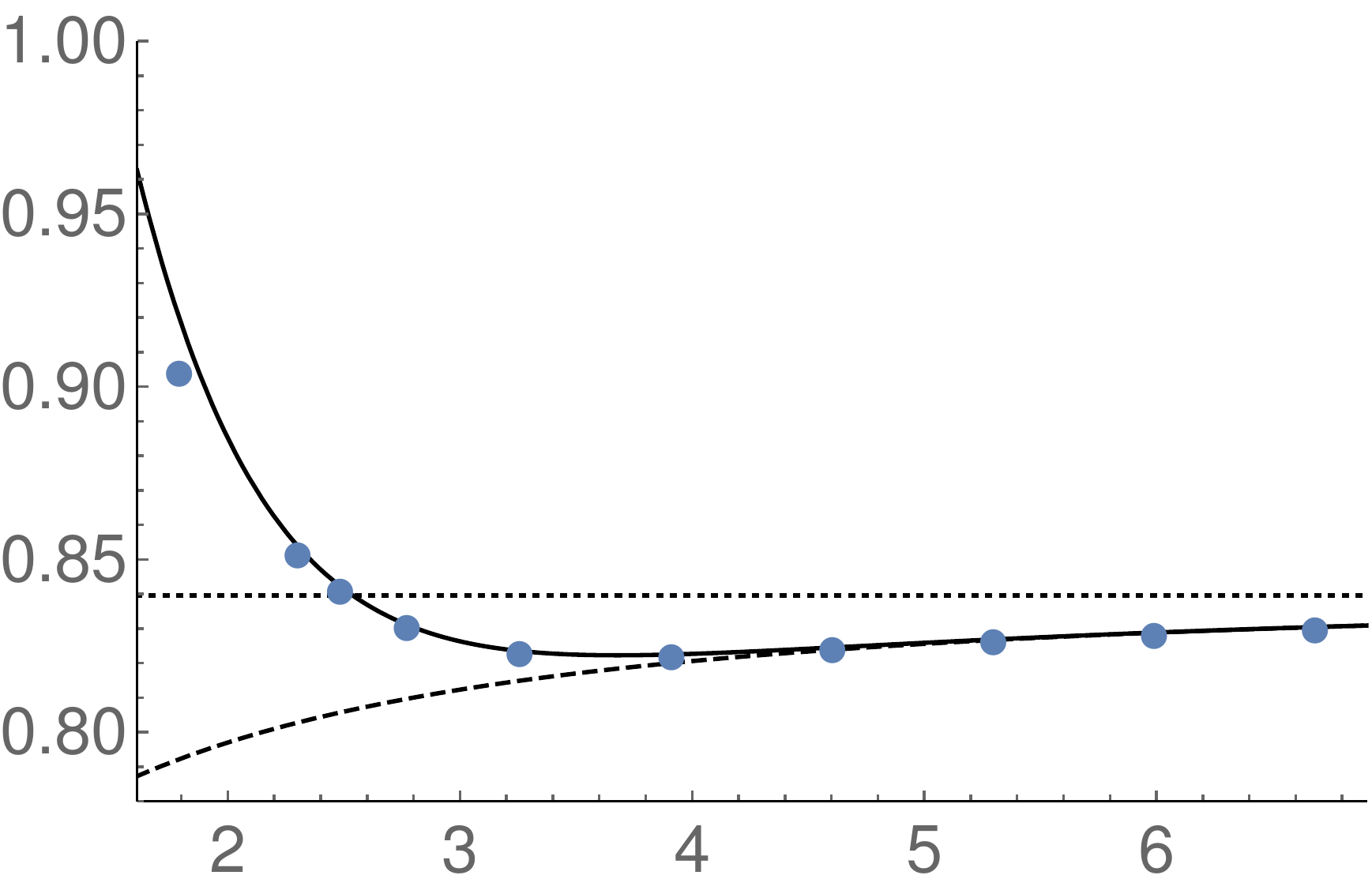}};
\end{tikzpicture}
}
\caption{\small The points represent numerical data for $\delta E_{{\rm odd}}\equiv E^{(+)}-E^{(-)}$ 
and $\delta E_{{\rm even}}\equiv E^{(+)}+E^{(-)}-\, v_{\rm F}\, e_\infty L$, which was computed
using the Bethe roots for the  state flowing to the conformal
descendent with dimensions $(\bar{\Delta},\Delta+1)$. The typical pattern
of roots for this state is given in fig.\,\ref{fig70}. On the left panel, the solid line is the prediction
coming from the finite size corrections eq.\,\eqref{Edifeq1}. On the right panel, 
the asymptotic formula \eqref{asympeq1a} with only the $L^{-1}$ term included
is displayed by the dashed line, 
while the solid line takes into account the $L^{-3}$ correction.
The limiting value $\lim_{L\to\infty}\frac{L}{2\pi v_{\rm F}}\,\delta E_{{\rm even}}=0.839583\ldots$ is shown by the dotted line.
Finally $s(L)$, which is needed to compute the r.h.s. of eqs.\,\eqref{Edifeq1},\,\eqref{asympeq1a}, was calculated
using the quantization condition \eqref{quantC1} with $\delta=\delta_+$ \eqref{deltaLvl1}.  The parameters were taken to be
$S^z=0$, $m=0$, ${\tt w}=0$, $\gamma=\frac{\pi}{5}$ and ${\tt k}=\frac{1}{20}$.
\label{fig90}}
\end{figure}
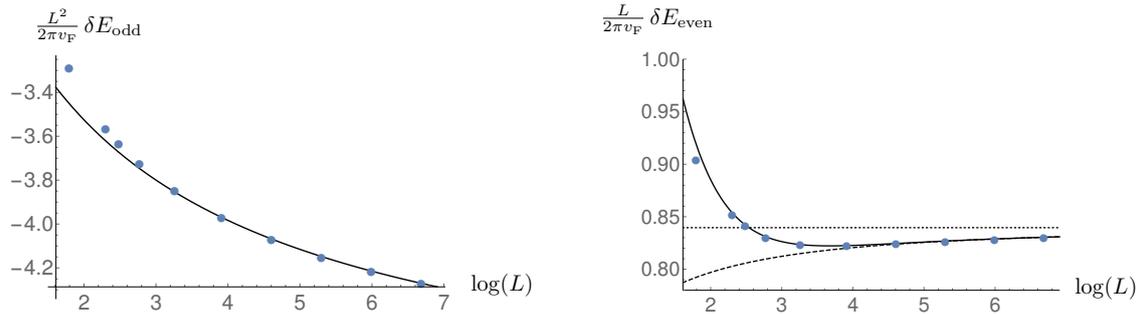

\bigskip
We are now in a position to discuss the finite size corrections
 to the energy spectrum of the alternating spin chain. 
 The results of our analysis suggest that its critical behaviour
 is described by a certain CFT whose Hilbert space is built out of
  the irreducible   (for general $n>0$)  representations 
  $\bar{{\cal V}}_{s,\bar{p}}\otimes{\cal V}_{s,p} $
of the left and right chiral $W_\infty$-algebras. As usual,
the finite size corrections are captured by irrelevant perturbations
of the critical Hamiltonian. In the case under consideration, 
these are strongly constrained due to the integrability of the spin chain with 
arbitrary length $L$. A useful property of the Hamiltonian \eqref{iasusay}, which follows immediately 
from eq.\,\eqref{Heq1}, is that it 
can be written in the form
\be
\mathbb{H}=\mathbb{H}^{(+)}+\mathbb{H}^{(-)}\,.\nonumber
\ee
The operators $\mathbb{H}^{(\pm)}$ are related through the $\mathbb{Z}_2$-transformation
${\cal D}$ \eqref{Deq2}
\be
\mathbb{H}^{(\mp)}={\cal D}\,\mathbb{H}^{(\pm)}{\cal D}\,,\nonumber
\ee
and read explicitly as
\bea\label{iaasuy}
\mathbb{H}^{(+)}&=&\frac{1}{2}\,\tan(\gamma)\sum_{j=1}^{2L}\,\sigma^z_j\,\sigma^z_{j+1}
-\frac{1}{\sin(2\gamma)}\,\sum_{j=1}^L\,\Big(\sigma^x_{2j}\,\sigma^x_{2j+2}
+\sigma^y_{2j}\,\sigma^y_{2j+2}+\sigma^z_{2j}\,\sigma^z_{2j+2}\Big)\nonumber\\[0.2cm]
&-&\frac{\ri}{2\cos(\gamma)}\,\sum_{j=1}^L
\Big(\,\big(\sigma^x_{2j}\,\sigma^y_{2j+1}-\sigma^y_{2j}\,\sigma^x_{2j+1}\big)\,
\sigma^z_{2j+2}+
\sigma^z_{2j}\,
\big(\sigma^x_{2j+1}\,\sigma^y_{2j+2}-\sigma^y_{2j+1}\,\sigma^x_{2j+2}\big)\nonumber
\Big)\nonumber\\[0.2cm]
&+&L\,\cot(2\gamma)
\eea
and
\bea\label{amsshsah}
\mathbb{H}^{(-)}&=&\frac{1}{2}\,\tan(\gamma)\sum_{j=1}^{2L}\,\sigma^z_j\,\sigma^z_{j+1}
-\frac{1}{\sin(2\gamma)}\,\sum_{j=1}^L\,\Big(\sigma^x_{2j-1}\,\sigma^x_{2j+1}
+\sigma^y_{2j-1}\,\sigma^y_{2j+1}+\sigma^z_{2j-1}\,\sigma^z_{2j+1}\Big)\nonumber \\[0.2cm]
&+&\frac{\ri}{2\cos(\gamma)}\,\sum_{j=1}^L
\Big(\,\big(\sigma^x_{2j-1}\,\sigma^y_{2j}-\sigma^y_{2j-1}\,\sigma^x_{2j}\big)\,
\sigma^z_{2j+1}+
\sigma^z_{2j-1}\,
\big(\sigma^x_{2j}\,\sigma^y_{2j+1}-\sigma^y_{2j}\,\sigma^x_{2j+1}\big)\nonumber
\Big)\nonumber \\[0.2cm]
&+&L\,\cot(2\gamma)\ .
\eea
All three operators commute and can be diagonalized simultaneously.
The eigenvalues $E^{(\pm)}$ of $\mathbb{H}^{(\pm)}$ are expressed in terms of
the solution to the BA equations as
\bea
{ E}^{(+)}=-\sum_{j=1}^M\frac{2\ri\sin(\gamma)}{\sinh(2\beta_j)+\ri\cos(\gamma)}\,, \qquad 
{ E}^{(-)}=\sum_{j=1}^M\frac{2\ri\sin(\gamma)}{\sinh(2\beta_{j})-\ri\cos(\gamma)}\ .
\eea
Contrary to the Hamiltonian $\mathbb{H}$, 
the difference $\mathbb{H}^{(+)}-\mathbb{H}^{(-)}$ is odd w.r.t. the transformation ${\cal D}$. 
This imposes heavy restrictions on the finite size corrections to $E^{(+)}-E^{(-)}$ for the Bethe states
of the form \eqref{state1}. 
Namely, the leading large-$L$ behaviour is described by the formula
\bea\label{Edifeq1}
\frac{{ E}^{(+)}-{ E}^{(-)}}{v_{\rm F}} =-\frac{2\pi \ri}{L^2}\ g_2\,\big(I_2^{(N)}-\bar{I}_2^{(\bar{N})}\big)+
o\big(L^{-2}\big)
\eea
with $I_2^{(N)}$ from eqs.\,\eqref{eigeq2},\,\eqref{iasussau} and the Fermi velocity is given by \eqref{Vfermi}.
The constant $g_2$ can be computed along the lines of  \cite{Lukyanov:1997wq} and reads explicitly as
\be
 g_2=
\frac{\sqrt{\pi}\,\Gamma(\frac{5}{2}+\frac{2}{n})\Gamma^2(1+\frac{1}{n})}{3\Gamma(\frac{2}{n})
\Gamma^2(\frac{3}{2}+\frac{1}{n})}\, .\nonumber
\ee
In practice eq.\,\eqref{Edifeq1} is useful for determining the scaling counterpart, i.e., the r.h.s. of eq.\,\eqref{state1}, 
for the Bethe state $|\psi\rangle$ even on the
lattice of size $L\sim 10$, where all three $2^{2L}\times 2^{2L}$-matrices $\mathbb{H}$, $\mathbb{H}^{(\pm)}$ 
can be diagonalized directly. On the left panel of fig.\ref{fig90}, we illustrate formula \eqref{Edifeq1}  for the state that as $L\to\infty$ flows 
to the
conformal descendent with dimensions $(\bar{\Delta},\Delta+1)$.

\bigskip

It is possible to extract the finite size corrections for $E={ E}^{(+)}+{ E}^{(-)}$
from the results of the work \cite{Lukyanov:1997wq}. In particular, one has that for $n>1$ 
\bea\label{asympeq1a}
\frac{{ E}^{(+)}+{ E}^{(-)}}{v_{\rm F}}&=&\,
e_{\infty} \,L+
\frac{2\pi }{nL}\ \Big(
I_1^{(N)}+\,\bar{I}_1^{(\bar{N})}\,\Big) \\[0.2cm]
&-&
\frac{2\pi}{L^3}\ \Big(\, 2\pi^2\,g_1 I_1^{(N)}\,\bar{I}_1^{(\bar{N})}+g_3 \,
 \big(\, { I}_3^{(N)}+
 {{ \bar{I}}}_3^{(\bar{N})}\, \big)\, \Big)+O\big(L^{-5},L^{-2n-1}\big)\ .\nonumber
\eea
Again the eigenvalues of the local IM $I_1^{(N)}$, $I_3^{(N)}$ are given in eqs.\,\eqref{eigeq2},\,\eqref{iasussau}.
The specific bulk energy $e_{\infty}$ reads as
\be
e_\infty=- \frac{4}{\pi}\int_{0}^\infty\rd t\ \frac{\sinh(\frac{2 t}{n})
}{\sinh(\frac{(n+2)t}{n })
\cosh(t)}\ ,\nonumber
\ee
while the value of the constants are
\be
g_1=-\frac{\cot(\frac{\pi}{n})}{2\pi\,n^2} \,, \ \ \ \ \ \ \ \ \ \ \ 
g_3= \frac{\pi \Gamma(\frac{7}{2}+\frac{3}{n})\Gamma^3(1+\frac{1}{n})}
{18 \Gamma(\frac{3}{n})\Gamma^3(\frac{3}{2}+\frac{1}{n})}\ .\nonumber
\ee
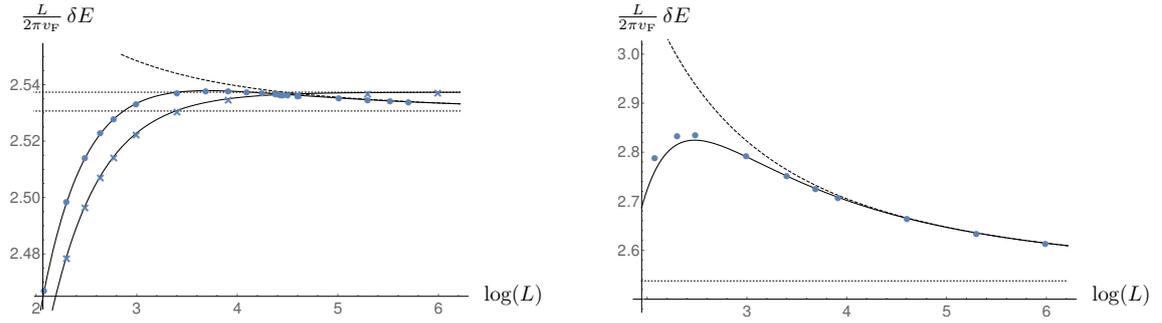
\begin{figure}
\centering
\scalebox{0.8}{
\begin{tikzpicture}
\node at (-3,2.7) {\small$\frac{L}{2\pi v_{\rm F}}\,\delta E$};
\node at (6.8,2.7) {\small$\frac{L}{2\pi v_{\rm F}}\,\delta E$};
\node at (4.5,-1.9) {\small $\log(L)$};
\node at (14.5,-1.9) {\small $\log(L)$};
\node at (0,0) {\includegraphics[width=0.5\textwidth]{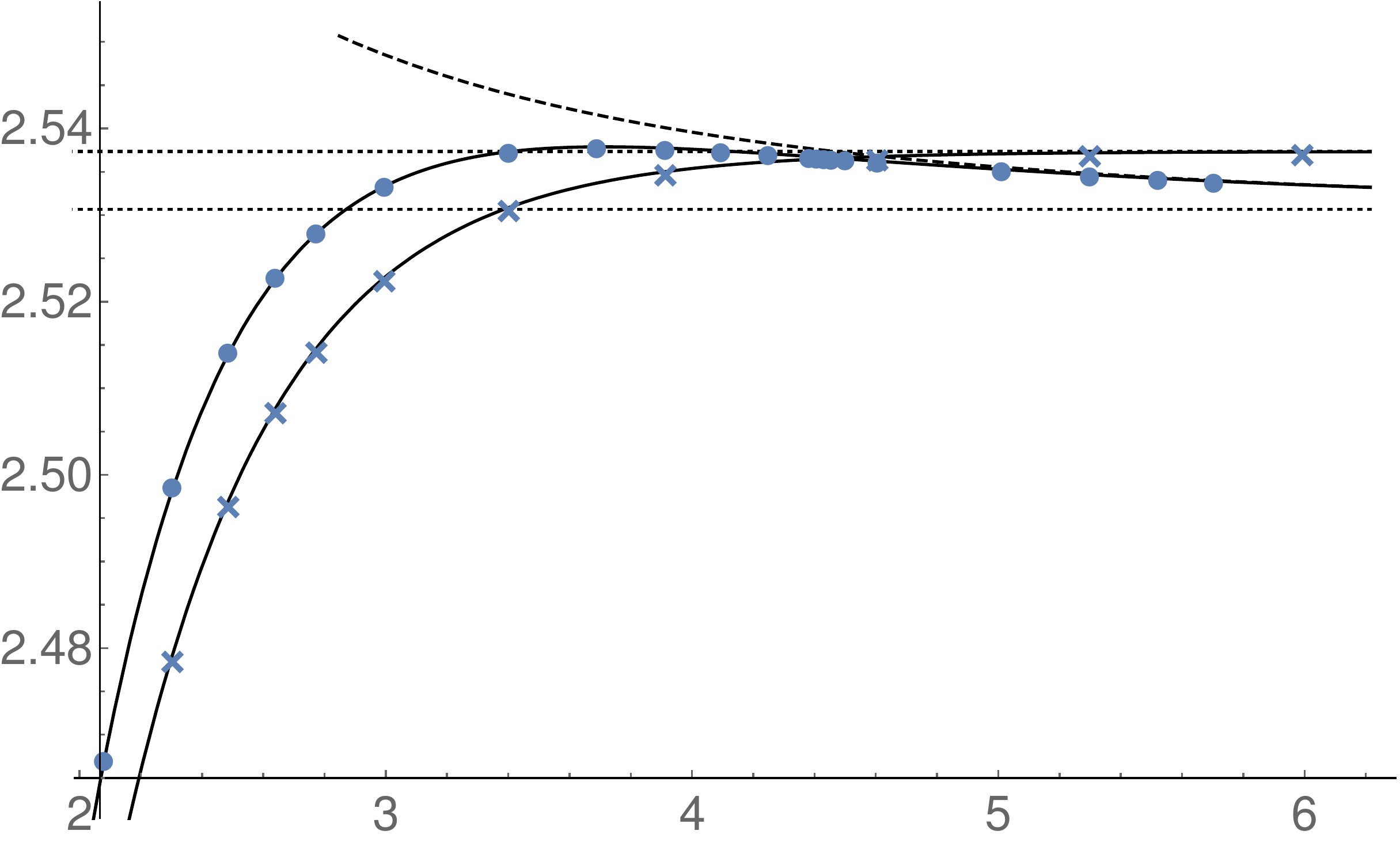}};
\node at (10,0) {\includegraphics[width=0.5\textwidth]{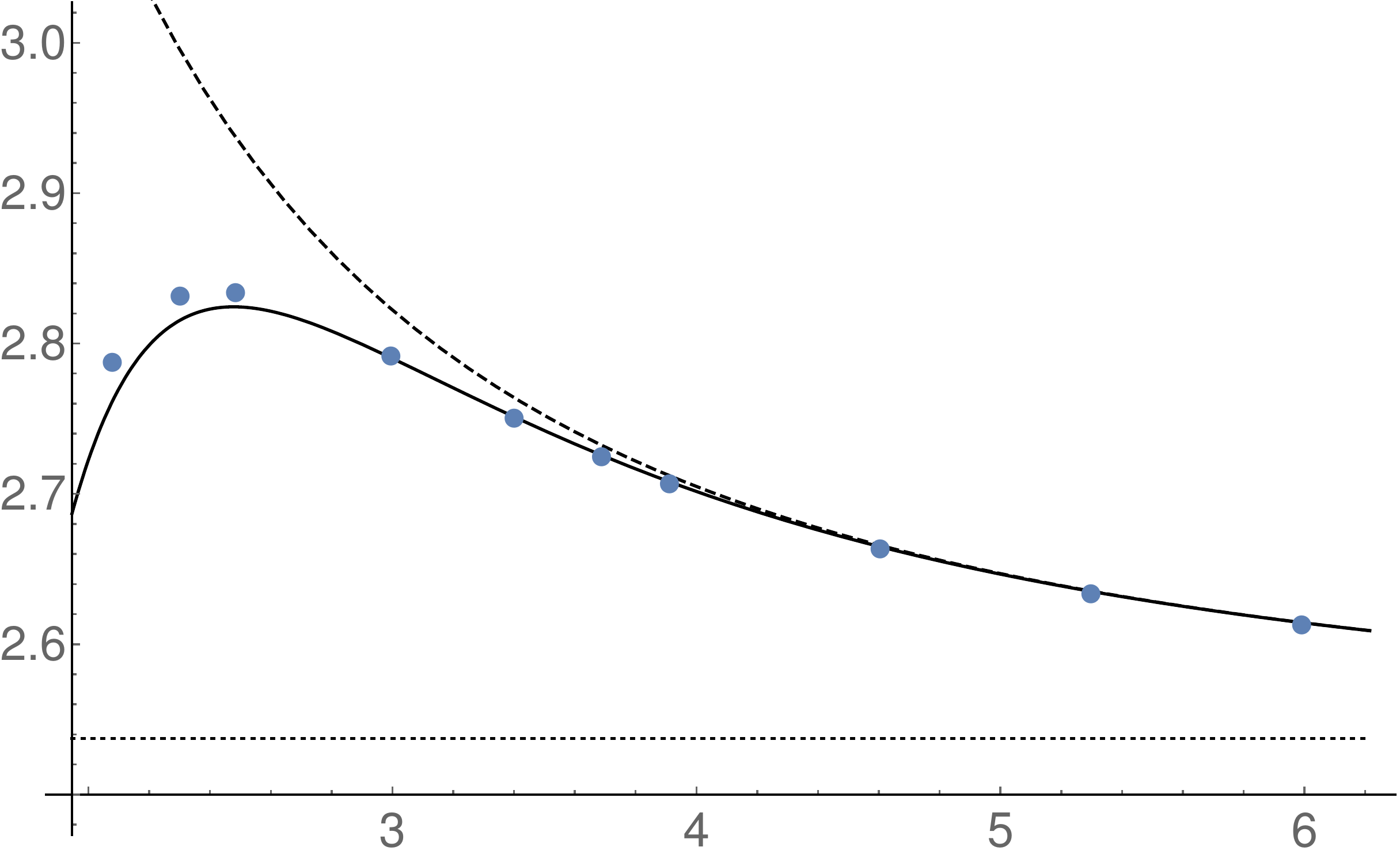}};
\end{tikzpicture}
}
\caption{\small Numerical data  obtained from the 
solution to the BA equations for the  winding states with ${\tt w}=1$, $S^z=0$, $m=0$ (left panel) and $m=2$ (right panel) is depicted 
for $\delta E\equiv E-\, v_{\rm F}\,e_\infty  L$.
On the left panel the crosses and circles correspond to the RG trajectories 
whose representatives for $L=20$
are shown on the left and right of fig.\,\ref{fig40}, respectively. 
On the right panel, the points correspond to the state whose typical pattern of Bethe roots is displayed in fig.\,\ref{fig60}.
Note that $L$ ranges from $L=8$ to $L=1000$.
In both panels the dashed lines come from the asymptotic formula eq.\,\eqref{asympeq1a} taking account only the $L^{-1}$ terms,
while to compute the solid lines the $L^{-3}$ correction was included.
The dotted horizontal lines show the limiting values of $\frac{L}{2\pi v_{\rm F}}\,\delta E$.
The value of $s(L)$ entering into the r.h.s. of eq.\,\eqref{asympeq1a} was calculated using the quantization condition \eqref{quantC1}
with $m=0$ (left panel) and $m=2$ (right panel). 
The parameters are set to be the same as in fig.\,\ref{fig40}, i.e., $\gamma =\frac{\pi}{5}$ and ${\tt k}=\frac{1}{25}$.
\label{fig100}}
\end{figure}
The formula \eqref{asympeq1a} calls for some important remarks. 
Firstly, with regards to the symbol $O\big(L^{-5},L^{-2n-1}\big)$, 
by this notation we mean that
\bea\label{susauas}
O(L^{-a},L^{-b})\equiv o(L^{-c})\ \ \  \ \ \ {\rm where}\ \ \  \ c={\rm min}(a, b)-\epsilon\ \ \ \ \  \ \  \forall \  \epsilon>0\ .
\eea
Corrections to \eqref{asympeq1a} proportional to integer powers of $L$ are expressed in terms of the
eigenvalues of the local IM. The next correction of this kind will involve combinations of the eigenvalues 
of $\mathbb{I}_m$ and $\bar{\mathbb{I}}_m$ with $m\le 5$. 
The term $\propto L^{-2n-1}$ comes from the dual non-local  IM
whose eigenvalue appeared in eqs.\,\eqref{Apeq1},\,\eqref{Apeq2} where it was denoted by $\tilde{H}_1^{(\pm,N)}$.
For $0<n<1$ it becomes the dominant correction to the
scaling, hence the restriction $n>1$ to the applicability of eq.\,\eqref{asympeq1a}.
The same happens in the $XXZ$ spin chain  (see ref.\,\cite{Lukyanov:1997wq}).
The next remark concerns the finite size corrections to the quantization condition \eqref{quantC1}, which was
 denoted by $O\big((\log L)^{-\infty}\big)$. In fact, we have numerically checked that
it goes as $O(L^{-2},L^{-2n})$ (see, e.g., tab.\,\ref{tab1}). It is easy to see that such an uncertainty in
``$s$'' will interfere with the subleading corrections presented in \eqref{asympeq1a}. At first glance,
this makes the $L^{-3}$ correction useless. However, in our numerical studies we found that formula \eqref{asympeq1a}
gives remarkably accurate results not only when ``$s$'' is zero identically, but also for non-zero $s$ 
satisfying the quantization condition with $m=0$ (see right panel of  fig.\,\ref{fig90} and left panel of fig.\,\ref{fig100}).
Note that for $m$ different from zero eq.\,\eqref{asympeq1a}  sometimes gives reasonable results as shown on the right panel of fig.\,\ref{fig100}, where
$m=2$. Unfortunately it is difficult to say \emph{a priori} when the $L^{-3}$ term improves the accuracy of the leading asymptotic behaviour.

\bigskip

It should be mentioned that the quantization condition
\eqref{quantC1},\,\eqref{asdasd1c},  the formula for $\Pi(L)$ \eqref{omegaeq1},\,\eqref{asdasd1c}, and 
the scaling of the energy \eqref{asympeq1a}, which are the main tools enabling a detailed investigation of the RG flow, can
be written in an elegant and compact way. To this end introduce the notation
\bea\label{hasysaysa}
{\mathfrak Y}(L)=
\bigg(
{\frac{1}{6}-\frac{2s^2}{n}-\frac{p^2+{\bar p}^2}{n+2}-N-{\bar N}}\bigg)  \log(L)+{\mathfrak Y}_0\ ,
\eea
where
\bea\label{iisasias}
{\mathfrak Y}_0=
\bigg(
{\frac{1}{6}-\frac{2s^2}{n}-\frac{p^2+{\bar p}^2}{n+2}-N-{\bar N}}\bigg)  \log\bigg(\frac{2\Gamma(\frac{3}{2}+\frac{1}{n})}{\sqrt{\pi}\Gamma(1+\frac{1}{n})}\bigg)
+
\frac{1}{n}\, \big(G^{(N)}+{\bar G}^{({\bar N})}\big)\ ,
\eea
and
$G^{(N)}$,
${\bar G}^{({\bar N})}$ were defined before  through eqs.\eqref{iasus1},\eqref{iasus2}.
It is easy to see that the quantization condition takes the form
\bea\label{ioasisis}
-2n\, \bigg(\frac{\partial {\mathfrak Y}}{\partial s}\bigg)_{{\tt k},L}= 2\pi m
+O(L^{-2},L^{-2n})\ \ \ \ \ 
\qquad  (m=0,\pm 1,\pm2,\ldots)\ ,
\eea
while
\be\label{Pieq2}
\Pi(L) =\exp\Bigg[
\frac{2n }{n+2}\ \bigg(\frac{\partial 
{\mathfrak Y}}{\partial {\tt k}}\bigg)_{s,L}+O(L^{-2},L^{-2n})\,\Bigg]
\ee
and the equation for the energy \eqref{asympeq1a}  is expressed as
\be\label{aisisai}
\frac{E}{v_{\rm F}}-e_\infty \,L=-2\pi\ \bigg(\frac{\partial 
{\mathfrak Y}}{\partial L}\bigg)_{s,{\tt k}}+O(L^{-3},L^{-2n-1})\ . 
\ee
The above relations point to the fundamental   r$\hat{\rm{o}}$le of the quantity ${\mathfrak Y}$.
In the following we will try to elucidate its meaning.

\bigskip
As it was first observed by Yang and Yang for the XXZ spin chain \cite{Yang1,Yang2}, the  BA  equations are obtained from a
certain variational principle. 
 In the case under consideration the  latter is formulated as follows.
Introduce  the Yang-Yang functional defined for an arbitrary set of complex numbers $\{\beta_j\}$
\bea\label{Ydef1}
Y\big[\{\beta_j\}\big]=\frac{n+2}{ n}\ \bigg[\,2\sum_{j}\big(L\,v(\beta_j)+({\tt k}-I_j)\,\big(\beta_j-\tfrac{\ri\pi}{4}\big)\,\big)+\sum_{j,m}u(\beta_j-\beta_m)\,\bigg]\ ,
\eea
where the 
functions $u(\beta)$ and $v(\beta)$ are related to $\Theta(\beta)$ and $P(\beta)$   in \eqref{func1} as
\bea\label{kasisai}
\Theta(\beta)=2\pi\ \partial_\beta u(\beta)\ ,\ \ \ \  \ P(\beta)=2\pi\ \partial_\beta v(\beta)\ .
\eea
The  BA equations  \eqref{klkwqnmsd} are equivalent to the condition for a local extremum
\bea
\delta Y\big[\{\beta_j\}\big]=0\ ,\nonumber
\eea
where all the parameters, including the set of the Bethe numbers $\{I_j\}$,
 are  given and kept  fixed under the infinitesimal variation  of  $\beta_j$.
Suppose now we have  a set  $ \{\beta_j\}$ solving 
the extremum condition, and consider the value of the Yang functional calculated on it.  
This
on-shell value will be denoted below by  $Y(L)$
emphasizing  its dependence on the length of the spin chain. 
As it is sensitive to the shift $\beta_j\to\beta_j+\ri\pi$ of any one of the Bethe
roots, 
to define $Y(L)$ unambiguously we assume that $0\le\Im m(\beta_j)<\pi$.
Notice that eq.\eqref{kasisai} specifies the functions $u(\beta)$ and $v(\beta)$  up to  additive constants,
which have no effect on the
extremum condition, but
the on-shell value of the Yang functional depends on them. It will be convenient for us to define
$u(\beta)$ as
\bea\label{aiaiasi}
u(\beta)
=\frac{\ri }{4\pi}\,\bigg(2\beta^2-\frac{\pi^2}{3}+\gamma(\pi-2\gamma)
+\rm{Li}_2\big(\re^{2\beta-2\ri\gamma}\big)
+\rm{Li}_2\big(\re^{-2\beta-2\ri\gamma}\big)
\bigg)\ .
\eea
Here ${\rm{Li}}_2(z)$ is a dilogarithm function, with the standard choice of branch cut 
 along the positive real axis $(1,+\infty)$ so that  $u(\beta)$ 
 is a single valued function in the complex plane $\beta$
 with the systems of cuts shown in the right  panel of fig.\,\ref{branch1}. Next, we specify   the function $v(\beta)$  by
 the formula
 \bea\label{vdef1}
 v(\beta)=-\frac{1}{4\pi}\ \Im m\Big({ \rm{Li}}_2\big(-\re^{2\ri\gamma}\big)\Big)+
\int_{0}^\beta \frac{\rd\beta'}{2\pi}\,  P(\beta')\ ,
 \eea
where the integration is taken along a straight line segment in the complex plane that connects the origin with the end point
$\beta$.
Thus defined, $v(\beta)$ is a single   valued  analytic function in the complex $\beta$-plane with two branch cuts
starting from the points $+\frac{\pi-2\gamma}{4}\,\ri$ and $-\frac{\pi-2\gamma}{4}\,\ri$ which extend along the imaginary axis to
$+\ri\,\infty$ and $-\ri\,\infty$, respectively. With this prescription there  is an ambiguity  in the computation
of $v(\beta_j)$ for the Bethe roots lying exactly on the imaginary axis with $|\Im m(\beta_j)|\geq \frac{\pi-2\gamma}{4}$. However 
for a general value of the twist parameter ${\tt k}\not=0$,  this subtlety can be ignored.

\bigskip
Consider the large-$L$ behaviour of  $Y(L)$, i.e., the on-shell value of the Yang functional. It is not difficult to see that it diverges 
quadratically as $L\to \infty$,
\bea\label{Ydiveq1}
Y(L)=Y^{\rm (div)}(L)+O\big(\log(L)\big) \nonumber
\eea
with the divergent part involving  the quadratic and linear terms
\bea\label{diveq2}
Y^{\rm (div)}(L)=y_\infty\, \frac{L^2}{4}+\frac{\ri\pi}{n}\ 
{\cal N}\, L\ .
\eea
Notice that $Y^{({\rm div})}(L)$ is somewhat non-universal. In particular,  
$y_\infty$ depends on our voluntaristic choice of the integration constants 
 appearing in the definition of $u(\beta)$ and $v(\beta)$. 
 Specifying these functions as in \eqref{aiaiasi},\,\eqref{vdef1}, it is straightforward to show that  \cite{Lukyanov:2011wd}
 \bea
y_\infty=-\int_{0}^\infty\frac{\rd t}{t^2}\ \bigg(\frac{\sinh(\frac{2t}{n})}
{\sinh(\frac{ (n+2) t}{n}) \cosh(t )}-\frac{2}{n+2}\bigg)\ .\nonumber
\eea
It turns out that for such $u$ and $v$ the linear divergent part in \eqref{diveq2} is pure imaginary
with ${\cal N}$ an integer.  
The latter is sensitive to local deformations of the branch cuts for the function $v$ and hence is not
particularly interesting.
We can ignore this term by focusing on 
\be
\exp\big(2n\,Y^{({\rm reg})}(L)\big)=\exp\big(2n (Y(L)-Y^{\rm (div)}(L))\big)
\ee
instead of the Yang functional itself. 
%
%
%
%
%
There exists a
remarkable relation between the quantity $\exp\big(2n\,Y^{({\rm reg})}(L)\big)$ and ${\mathfrak Y}(L)$ introduced
 in eq.\eqref{hasysaysa}. Namely
\bea\label{asiasiasq}
\exp\big( 2n\,Y^{\rm (reg)}(L)\big)=
\re^{ 2\pi\ri \, (n+2)\, ({\tt k}\,{\cal N}_1+2\,{\cal N}_2)}\ 
\exp\Big(2n\,{\mathfrak Y}(L)+2\pi m s
+O(L^{-2},L^{-2n})\Big)\ .
\eea
Here the variable $s$ is defined by \eqref{Beq1},\,\eqref{seq1}
and is related to the integer $m$ through the quantization condition\ \eqref{ioasisis},
while ${\cal N}_1$ and ${\cal N}_2$ appearing in the phase factor are integers.
The value of these last two depends on the specification of the branch of the
multi-valued function 
$\mathfrak{Y}_0$ \eqref{iisasias} entering the r.h.s. of eq.\,\eqref{asiasiasq} via $\mathfrak{Y}(L)$. 
Note that as it follows from the formula \eqref{Pieq2} for $\Pi(L)\equiv\prod_j\re^{4\beta_j}$ 
and eq.\,\eqref{asiasiasq}
one has
\be
\prod_{j=1}^M\,\re^{2(\beta_j-\frac{\ri\pi}{4})}=(-1)^{{\cal N}_1}\,\exp\Bigg[
\frac{n }{n+2}\ \bigg(\frac{\partial 
{\mathfrak Y}}{\partial {\tt k}}\bigg)_{s,L}+O(L^{-2},L^{-2n})\,\Bigg]
\ee
(recall that in this paper $L$ and $M$ are assumed to be even). Also our numerical work shows
 that there is no simple relation between the vanishing as
$L\to\infty$ corrections 
$O(L^{-2},L^{-2n})$ for $Y(L)$ and the  corrections for the energy $E(L)$ indicated in \eqref{aisisai}.
Thus it  is unlikely that there is a simple formula  that systematically relates the large-$L$ asymptotic expansions  of
 $Y(L)$ and $E(L)$.
\bigskip

The following comments are in order here.
In the case  of the ground states, with zero $m$ and ${\tt w}$, the scaling function
 ${\mathfrak Y}(L)$ was calculated in  \cite{Lukyanov:2011wd}. Those results allows us to specify the $s$- and $p$-independent constant
 in formula \eqref{iasus2}  which affects eqs.\eqref{hasysaysa},\,\eqref{iisasias} and hence, is important in the derivation of
 \eqref{asiasiasq}.
Also the special r$\hat{\rm{o}}$le of the Yang functional for finding the spectrum
in a  quantum integrable system
 was emphasized by Nekrasov and Shatashvili \cite{Nekrasov:2009rc}. The quantization condition for $s$
studied here provides  an illustration of the general phenomena.

\bigskip

To finish, let us underline  the main result  reported in this work.  
 We studied the
integrable spin chain whose critical behaviour is governed by a CFT possessing a continuous spectrum of scaling dimensions.
In such a situation  one of the first questions that needs to be addressed concerns the density of states of the continuous theory.
Using the powerful method of the ODE/IQFT correspondence we  determined the  phase shifts, which appear in the
 quantization condition for the spectrum 
of the chain for large but finite lattice size $L$. As usual, the phase shifts are simply related to the density of states of
the continuous theory.
 In the remarkable  work \cite{Ikhlef:2011ay}, it was observed 
that in the scaling limit,
the density of  ``primary'' Bethe states of the spin chain  coincides with the  density of states in
the $SL(2,R)/U(1)$ black hole. This observation lead the authors of  \cite{Ikhlef:2011ay} to conjecture that
the latter  describes the universal scaling behaviour of the spin chain. Our work was mainly  motivated by this interesting
hypothesis but, unfortunately, the obtained results  rule it out. 
We  demonstrated  that 
the density of the descendent states  of the spin chain in the scaling limit  is
not
what
 is expected for the $SL(2,R)/U(1)$ black hole \cite{Maldacena:2000hw,Maldacena:2000kv,Hanany:2002ev}.
Another important issue discussed in this work  is related to the presence of
  ``bound''
   states corresponding to pure imaginary values of $s$ in the CFT 
  governing the critical behaviour of the alternating spin chain.

\acknowledgments

The authors gratefully acknowledge support from the Simons Center 
for Geometry and Physics, Stony Brook University 
during the workshop 
``Exactly Solvable Models of Quantum Field Theory and Statistical Mechanics''
(September 4 -- November 30, 2018), where this work was initiated.
The authors thank Holger Frahm for stimulating discussions and correspondence.
We are also grateful to Hubert Saleur for reading 
the draft of the paper and for important comments.

\end{document}